\documentclass[singlecolumn, prb, showpacs, superscriptaddress]{revtex4-1}
\usepackage{graphicx}
\usepackage{dcolumn}
\usepackage{bm}
\usepackage{color}
\usepackage{amsmath}

\makeatletter
\AtBeginDocument{\@ifpackageloaded{natbib}{\ifNAT@numbers\if@filesw\immediate\write\@auxout{\string\global\string\NAT@numberstrue}\fi\fi}{}
\let\($\let\)$}
\makeatother

\begin{document}

\title{Fundamental insights to topological quantum materials: \\A real-space view of 13 cases by supersymmetry of valence bonds approach}

\author{F. C. Chou}
\affiliation{Center for Condensed Matter Sciences, National Taiwan University, Taipei 10617, Taiwan}
\affiliation{National Synchrotron Radiation Research Center, Hsinchu 30076, Taiwan}
\affiliation{Taiwan Consortium of Emergent Crystalline Materials, Ministry of Science and Technology, Taipei 10622, Taiwan}
\affiliation{Center of Atomic Initiative for New Materials, National Taiwan University, Taipei 10617, Taiwan}

\date{\today}

\begin{abstract}

We present a real-space view of one-dimensional (1D) to three-dimensional (3D) topological materials  with 13 representative samples selected from each class, including  1D \textit{trans}-polyacetylene, two-dimensional (2D) graphene, and 3D topological insulators, Dirac semimetals, Weyl semimetals, and nodal-line semimetals.  This review is not intended to present a complete up-to-date list of publications on topological materials, nor to provide a progress report on the theoretical concepts and experimental advances, but rather to focus on an analysis based on the valence-bond model  to help the readers gain a more balanced view of the real-space bonding electron characteristics at the molecular level versus the reciprocal-space band picture of  topological materials.  Starting from a brief review of low-dimensional magnetism with ``toy models'' for a 1D Heisenberg antiferromagnetic (HAF) chain,  1D \textit{trans}-polyacetylene and 2D graphene are found to have similar conjugated \(\pi\)-bond systems, and the Dirac cone is correlated to their unconventional 1D and 2D conduction mechanisms.  Strain-driven and symmetry-protected topological insulators are introduced from the perspective of material preparation and valence-electron sharing in  the valence-bond model analysis.  The valence-bond models for the newly developed  Dirac semimetals, Weyl semimetals, and nodal line semimetals are examined with more emphasis on the bond length and electron sharing, which is found consistent with the band picture.  The real-space valence-bond analysis of topological materials with a conjugated \(\pi\)-bond system suggests that these topological materials must be classified with concepts borrowed from group theory and topology, so that a supersymmetry may absorb the fluctuating broken symmetry. Restoration of a thermodynamic system with higher entropy (i.e., the lower Gibbs free energy) is more appropriate to describe such topological materials instead of the traditional material classification with the lowest enthalpy for the presumed rigid crystal structure.

\end{abstract}

\maketitle

\tableofcontents
\newpage
\setcounter{tocdepth}{2}

\section{General introduction}

A common theme of  Nobel prizes in chemistry and physics for the past two decades has been the discovery and interpretation of the unusual physical properties of condensed matter originating from reduced electronic dimensionality, including the 2000 chemistry award for one-dimensional (1D) conducting polymers,\cite{Shirakawa:2006en, FincherJr:1982hc} the 2010 physics award for two-dimensional (2D) graphene,\cite{Geim:2011ed, Novoselov:2011gq} and the 2016 physics award for topological quantum materials.\cite{Haldane:2017et}  Starting from the surprising discovery of 1D conducting polymers and the first isolation of graphene from graphite as a perfect mono-atomic 2D material, the study of ``toy models'' for the 1D Heisenberg spin chain system turns out to build the foundation for  newly developed topological materials having topologically nontrivial 2D electronic edge states in three-dimensional (3D) condensed matter.\cite{Hasan:2010kua}  

Since 2005, more and more topological materials have been predicted and categorized through the combination of theoretical prediction and angle-resolved photoemission spectroscopy (ARPES) experiments,\cite{Bansil:2016bu} and the search of topological materials has lately shifted to using existing materials databases with designed algorithms in a massive hunt for new topological materials.\cite{E:2018wp} The understanding of this new class of material has  mostly relied on their calculated band structure with added variables, including spin-orbit coupling (SOC) and strain, and the categorization of materials with terminology borrowed from the mathematics of topology and group theory, not to mention the increasing number of involved physical concepts introduced to address the emerging quantum phenomena from the theoretical perspective of the band picture, including the quantum spin Hall effect, topological order, Dirac cones, Chern invariants, Berry phase, fractional charge, and many more to come.  Clearly, the required intuitive interpretation from the real-space chemical-bond perspective is missing, which has created an unnecessary barrier for the materials scientists who focus on new material development and device applications. 

Here, we provide a brief historical review of topological material development from the chemical perspective with a simple valence-bond model interpretation and make a concise comparison between the similar concepts  with confusingly different names given by  chemists and physicists.  By using the selected representative topological materials with their calculated or ARPES-detected band pictures, a valence-bond symmetry analysis is provided to assist the reader to  intuitively understand the concept of ``topologically nontrivial'' categorization. This review is not intended to collect and list all topological materials reported to date following the path of theoretical physicists; instead, we pick representative topological materials from each major class in the mode of an experimental materials physicist who focuses on crystal growth and material design. A total of 13 samples are selected from the simplest 1D \textit{trans}-polyacetylene (\textit{trans}-PA), to 2D graphene, to 3D topological materials including topological insulators (TIs), topological crystalline insulators, topological Dirac semimetals, topological Weyl semimetals, and nodal-line semimetals. The crystal and electronic structures are analyzed with a proposed valence-bond model based on the symmetry of atomic coordination at the molecular level and the atomic electron configuration.  A comparative interpretation is provided for topological materials between the real-space valence-bond supersymmetry and the band picture in reciprocal space.  

\subsection{Brief review of topological materials}

The 2016 Nobel Prize in Physics went to David J. Thouless, F. Duncan, M. Haldane, and J. Michael Kosterlitz for their contribution to the understanding of topological quantum materials, mostly to recognize their important theoretical effort that paved the way to understand the recently verified ``topological states of matter'' (i.e., materials having entangled quantum states that can be transformed smoothly with protected symmetries).\cite{Haldane:2017et} The understanding and categorization of topological quantum materials has   relied heavily on the conceptual and mathematical approach via the topological analogy of band pictures.  Many such approaches involved physical concepts, including the Berry phase, Hall conductance, the quantum spin Hall effect, spin-orbit coupling, time-reversal symmetry, and \(Z_2\) topological invariants, were used to interpret and classify topological quantum materials.\cite{Bansil:2016bu, Hasan:2010kua}  In fact, the concept of topology has already been applied to categorize isomers for chemicals in real space in the field of chemistry,\cite{Hosoya:1975ea, Dobrowolski:2003tq} but  physicists focus more on the band picture of topological invariance.  A classic isomorphic first-order phase transition in the band picture was demonstrated by  Lifshitz in 1960 on metal under pressure and was named the ``Lifshitz transition.''\cite{Lifshitz:1960ux} 

The discovery of topological quantum materials is closely related to the effort made  to understand the low-dimensional materials that are 1D conducting polymers and  2D graphene, mostly because of the commonality of the reduced  electronic dimensionality due to the emerging types of valence-electron sharing, as reflected by the existence of conjugated \(\pi\)-bond systems in real space and the Dirac-cone-shaped band pictures found in graphene.\cite{Neto:2009wq}  We believe that the fundamental connection between these classes of material should be re-examined from the valence-bond perspective, which emphasizes   electronic dimensionality and  coordination at the molecular level. The 2D band picture of the Dirac-cone shape indicates a 2D conduction mechanism via massless Dirac fermion tunneling, but the increasing complexity of mixed bulk and surface bands for topological materials has reached such a level that a reliable analysis of experimental data  becomes difficult.  In addition, the topological orders in 2D have  been  mathematically described mostly in analogy to the quantum spin Hall effect of gapless edge states and the \(k\)-space band topology.  To understand the essential effect of topological indexing on topological material classification in real space, a reasonable valence-bond model is needed to serve as a complementary tool in  research. The topological order can be understood in real space by considering a valence-bond model that shows novel electron-sharing mechanisms of limited and degenerate configurations governed by a specific supersymmetry, so that the perturbation from local symmetry breaking due to Peierls condensation is absorbed (i.e., the instantaneous symmetry-breaking and restoration under quantum fluctuation is ``protected'' by a supersymmetry).  In this review, we  summarize major groups of 1D and 2D topological materials with representative 1D \textit{trans}-PA and 2D graphene. We first apply a more pedagogical manner, and many predicted or verified 3D topological materials are discussed by comparing their real-space valence-bond analysis and the reciprocal-space band picture.

\subsection{Valence-bond perspective of one-dimensional Heisenberg antiferromagnetic chain}

\begin{figure}
\begin{center}
\includegraphics[width=4.5in]{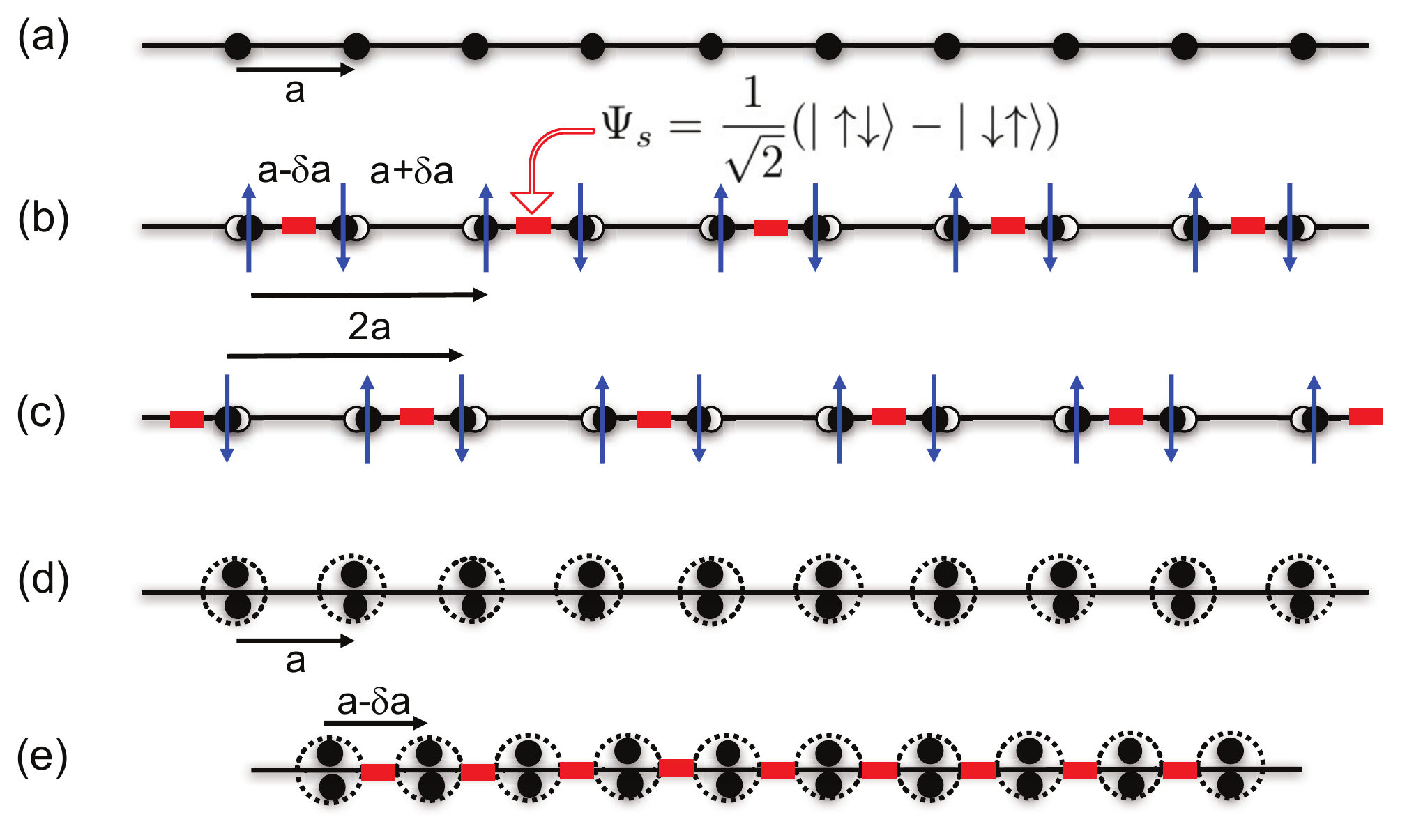}
\end{center}
\caption{\label{fig-chains} Toy models of 1D Heisenberg antiferromagnetic (HAF) chains. (a) 1D \(S=\frac12\) HAF spin chain or one valence electron per atom in a chain before the interaction is turned on. (b) A dimerized ground state of the \(S=\frac12\) HAF spin chain with the translational-symmetry breaking via periodicity change from \(a\) to \(2a\) induced by the spin-Peierls condensation, where valence bonds are formed at the singlet ground state with wave function \(\Psi_S\). (c) An alternative and degenerate valence-bond arrangement to that shown in panel (b).  (d) 1D \(S=1\) spin chain or two valence electrons per atom in triplet state before the interaction is turned on. (e) Valence-bond ground state of 1D \(S=1\) HAF spin chain with Peierls condensation without breaking the symmetry.}
\end{figure}

In the field of low-dimensional magnetism, the 1D Heisenberg spin chain has  served as a standard toy model for the study of quantum phenomenon (see Fig.~\ref{fig-chains}), mainly because the quantum fluctuation is most pronounced in low dimensions.  Bethe's pioneering work (the Bethe \(ansatz\)) made a breakthrough for solving the gapless ground state of a \(S=\frac12\) Heisenberg antiferromagnet (HAF) chain with nearest-neighbor coupling.\cite{Bethe:1931iz}   A dimerized \(S=\frac12\) HAF chain must break the translational symmetry as a result of antiferromagnetic (AF) spin pairing with concomitant Peierls condensation; i.e., a 1D spin-Peierls phase transition that opens a gap at the Brillouin zone boundary after the real-space periodicity is doubled from \(a\) to \(2a\), which could also be viewed as the origin of chemical bonding.  The 1D HAF chain solved by  Bethe's \(ansatz\) could be viewed as a resonating-valence-bond  state under quantum fluctuation to restore the translational symmetry broken by the dimerized condensation.\cite{Affleck:1987te}  Bonner and Fisher's attempt to solve the 1D HAF chain digitally has become the standard tool to estimate the spin-exchange coupling constant without  exactly solving the Hamiltonian.\cite{Bonner:1964fm}  With added next-nearest-neighbor coupling of \(J_{nnn}=\frac12J_{nn}\) to the \(S=\frac12\) HAF chain, Majumdar and Ghosh obtained two degenerate dimerized ground states (MG chain), as illustrated in Figs.~\ref{fig-chains}(b) and \ref{fig-chains}(c), under quantum fluctuation.\cite{Majumdar:1969iu}  Haldane conjectured that an integer HAF chain must have a spin gap in the limit of an infinitely long chain  (Haldane chain),\cite{Haldane:1983wc} and many \(S=1\) spin-chain compounds have been explored to support this claim.\cite{Vasilev:2005tf}  Affleck, Kennedy, Lieb, and Tasaki (AKLT) later used the concept of a valence bond [i.e., a spin pair in singlet form with wave function \(\Psi_S\)=\(\frac{1}{\sqrt{2}}(|\uparrow\downarrow\rangle-|\downarrow\uparrow\rangle\))] to solve for a unique ground state called a valence-bond solid  without broken symmetries for the Haldane chain, as illustrated in Fig.~\ref{fig-chains}(e). This shows an entangled Peierls condensate among atoms of \(S=1\).\cite{Affleck:1987te}  It was unexpected that playing with these simple toy models  in low-dimensional magnetism would pave the way to understanding both graphene and topological materials, as recalled by F. Haldane in his Nobel lecture.\cite{Haldane:2017et}          

Most low-dimensional materials classified as 1D or 2D have unavoidable interchain or interplane interactions as  condensed matter of higher dimension, so it is nearly impossible to identify compounds that can be viewed as a true realization of a MG- or a Haldane-chain system.  The successful isolation of graphene from graphite as a prototype mono-atomic layered 2D material is particularly inspiring,\cite{Geim:2011ed} mostly because of the close relationship between the conjugated \(\pi\)-bond system in real space and the observed Dirac-cone band picture also seen in most topological materials.\cite{Haldane:2017et, Hasan:2010kua}  Upon tracing the critical role played by the conjugated \(\pi\)-bond system  in 1D compounds, we find that  1D \textit{trans}-PA as a conducting polymer, which led to the Nobel prize in chemistry in 2000, could be a perfect real-life example of an 1D MG chain. This system provides the best bridge between the theory-experiment and the chemistry-physics divides.  Following the three milestone material classes awarded by the Nobel prize (i.e.,  1D conducting polymers,  2D graphene, and  3D topological quantum materials), we selected representative materials in three classes and use the valence-bond symmetry analysis to provide a consistent interpretation of why these materials are classified to show characteristics of unique conducting properties.  In particular, the real-space valence-bond symmetry and the band picture in reciprocal space are compared repeatedly  to test the consistency between the two views, so that the understanding of topological materials  no longer remains at the level of a conceptual analogy of topology. 

\section{Polyacetylene: one-dimensional topological material}

\subsection{One-dimensional \textit{trans}-polyacetylene as Majumdar--Ghosh chain}

\begin{figure}
\begin{center}
\includegraphics[width=5.0in]{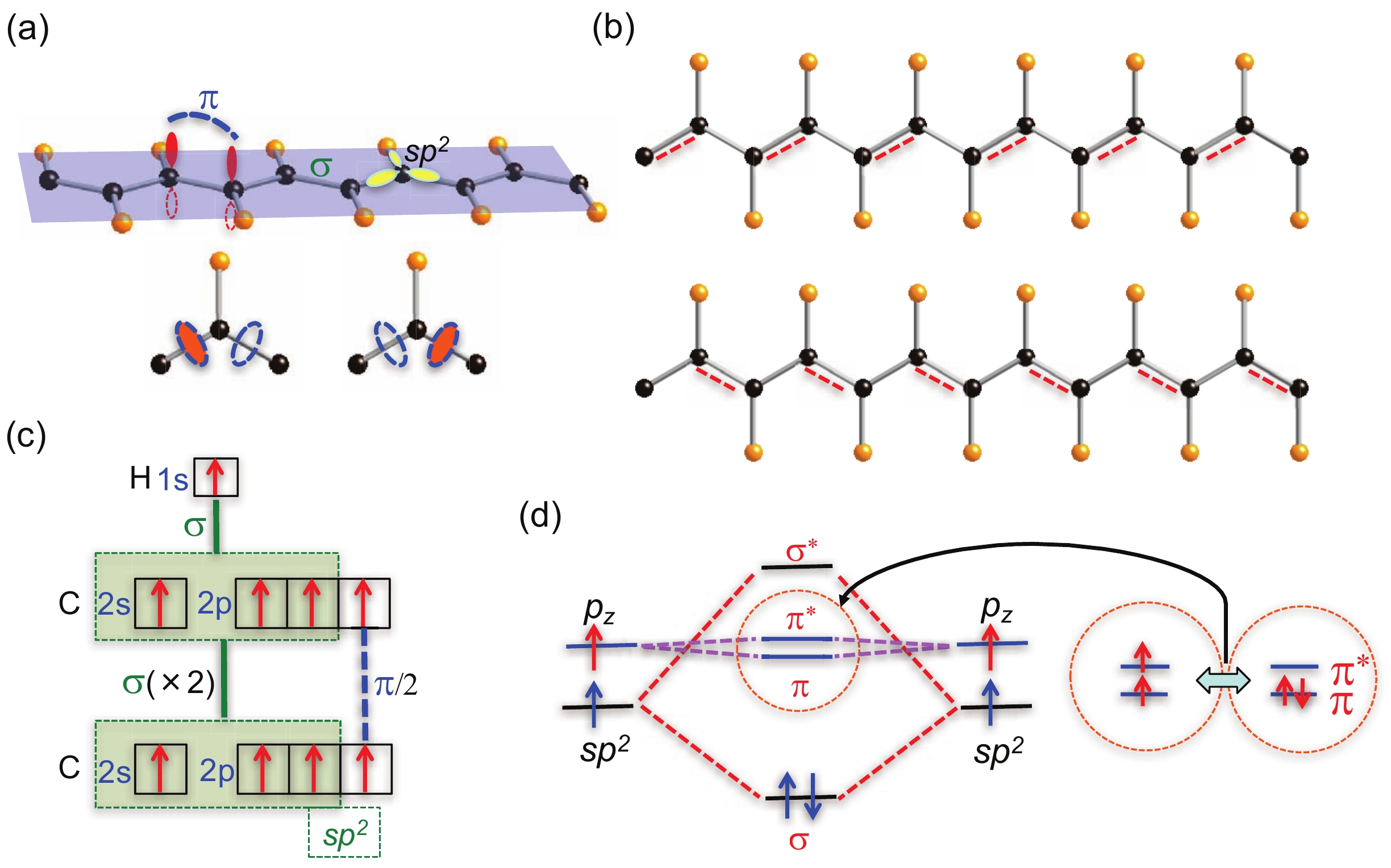}
\end{center}
\caption{\label{fig-PA} (a) Crystal structure of 1D \textit{trans}-polyacetylene (CH)\(_n\). Two \(\sigma\) bonds  form among two neighboring C--C via \(sp^2\) hybridized orbital overlap and a third \(\sigma\) bond forms  with H.  (b) Two degenerate configurations of broken symmetry as a result of \(\pi\)-bond-induced Peierls condensation, but the conjugated nature allows the distortion-restoration to be absorbed by a time-reversal symmetry, which is a real-life example of the theoretical MG chain, as shown in Figs.~\ref{fig-chains}(b) and \ref{fig-chains}(c). (c) Valence-bond model proposed for 1D \textit{trans}-PA. (d) The molecular orbital energy diagram for the \(\sigma\)- and \(\pi\)-bonding relationship, where the conjugated \(\pi\)-bond system is equivalent to the SOC mechanism of energy exchange between magnetic and phonon origins. }
\end{figure}

Although the \(S=\frac12\) HAF spin chain system has been tackled by physicists in the field of quantum magnetism, it seems difficult to identify a real compound that exactly fits the criteria shown in the Hamiltonian, such as a chain of atoms with one electron in the valence shell per atom to connect via the valence bond of electron pairs in  a spin-singlet state only.  However, we find that  1D \textit{trans}-PA with the chemical formula  (CH)\(_n\) may be the closest real-life example of a MG chain---provided that we focus only on the \(\pi\)-bond electrons  and leave out the \(\sigma\)-bond electrons which are responsible for the molecular structure of the zigzag chain coordination [see  Fig.~\ref{fig-PA}(a)].  In fact, without actually solving in detail for the ground state from the Hamiltonian, the variational method  implemented in the H\"uckel theory can satisfactorily handle the \(\pi\)-electron system with topological indexing for  conjugated hydrocarbon compounds.\cite{Hosoya:1975ea}      

Starting from the atomic electron configuration of carbon ([He]\(2s^22p^2\)), each carbon has a triangular coordination with two neighboring carbon atoms and one hydrogen in the (CH)\(_n\) chain, which suggests that \(\sigma\) bonds are formed among C-C and C-H pairs via orbital overlap of half-filled \(sp^2\) hybrid orbitals in \(\sigma\) bonds, as shown in Fig.~\ref{fig-PA}(c).  Considering that the unpaired electron in \(2p_z\) per carbon may form a weaker \(\pi\) bond with the neighboring \(2p_z\) of carbon via side-to-side orbital overlap and dimerize as a result of Peierls condensation, translational symmetry must be broken, as revealed by the periodicity change from \(a\) to \(2a\) [see Figs.~\ref{fig-chains}(b) and \ref{fig-chains}(c)], which is also described as bond-length alternation, with supporting experimental evidence.\cite{FincherJr:1982hc} 

Because there are two equivalent choices of \(\pi\)-bond formation for each carbon atom on either side of the neighboring carbon in a 1D \textit{trans}-PA chain, as shown in Fig.~\ref{fig-PA}, either configuration is implied to exist with a 50\(\%\) chance statistically under quantum or thermal fluctuation; and the configurations are entangled. The  \(\pi\) bond is weaker  than the \(\sigma\) bond because of the side-to-side orbital overlap of unpaired electrons in \(2p_z\) with much higher energy, as shown in Fig.~\ref{fig-PA}(d).  It is likely that the statistical choice of  the \(\pi\)-bond configuration is because it not only has local symmetry-breaking and restoration spontaneously as a conjugated system, but also because it is entangled throughout the whole chain in real space under quantum fluctuation.  Experimentally,  undoped 1D \textit{trans}-PA is gapped with identified bond alternation of broken translational symmetry and therefore falls into a nondegenerate dimerized ground state, and  doped 1D \textit{trans}-PA via intercalation has enhanced conductivity as a unique conducting polymer.  If the interchain van der Walls-type coupling is ignored completely, it is reasonable to view the role of hole doping being used to close the 1D band gap via breaking the permanently dimerized ground state and shifting the system into a chain having a conjugated \(\pi\)-bond system as a gapless ground state, which has been described in detail via concepts of soliton and fractional charge to explain why doped 1D \textit{trans}-PA shows the unexpected enhanced conductivity.\cite{BAERISWYL:1995ti, Kivelson:2001js}   In fact, 1D \textit{trans}-PA could be viewed as the first 1D topological phase identified, which has also been hinted by the analogy found in the Bi\(_2\)Se\(_3\)-In\(_x\)Bi\(_{2-x}\)Se\(_3\) superlattice built from the well-known 3D topological insulator Bi\(_2\)Se\(_3\).\cite{Belopolski:2017cp}

In a thermodynamic system, the existence of a conjugated \(\pi\)-bond system indicates a contribution of positive entropy change (\(\Delta S>0\)) moving from the dimerized nondegenerate ground state to the degenerate ground state of multiple configurations in fluctuation. That is, although it seems that the spontaneous \(\pi\)-bond breaking and restoration is unfavorable to stay at the lower enthalpy \(H\) of a thermodynamic system, it may become favorable when the enhanced entropy  reduces the Gibbs free energy of the thermodynamic system via  \(G=H-TS\).  Above all, the small and instantaneous distortion-restoration mechanism that temporarily breaks the translational symmetry  could be absorbed by a supersymmetry that contains a larger basis of three neighboring (spins) atoms in a group for 1D \textit{trans}-PA.  The supersymmetry can be identified to be the time-reversal symmetry, so that  a new gapless ground state of symmetry-protected topological order is established.

\subsection{Structure evolution from one-dimensional \textit{trans}-polyacetylene to two-dimensional graphene}

\begin{figure}
\begin{center}
\includegraphics[width=5.5in]{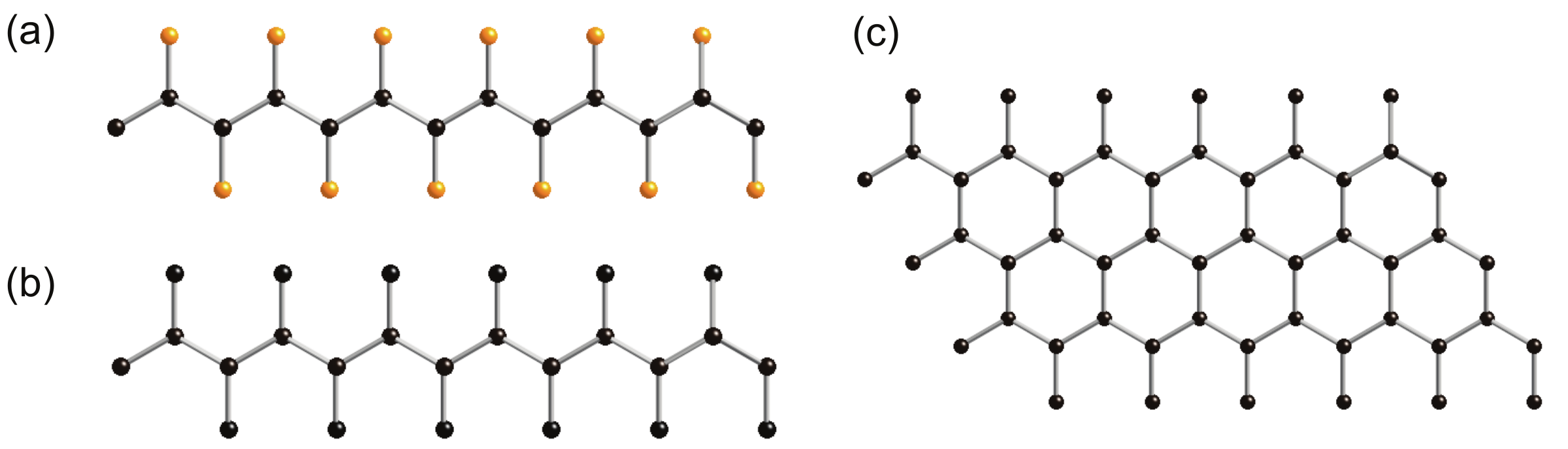}
\end{center}
\caption{\label{fig-PAtoGr} The similar electronic configuration and physical properties between 2D graphene (C\(_2\))\(_n\) and 1D \textit{trans}-PA  are expected from the evolution of dimensionality.  Starting from (a) 1D \textit{trans}-PA, we substitute H with C to form a virtual compound of 1D (C\(_2\))\(_n\), as shown in panel (b). Next, 2D graphene is constructed with 1D (C\(_2\))\(_n\) along the interchain direction, as shown in panel (c). }
\end{figure}

The crystal structure of graphene is well known for its honeycomb lattice composed of triangular coordinated carbon atoms.  It is interesting to note the close relationship between 1D \textit{trans}-PA and 2D graphene, which have identical carbon triangular coordination, both crystallographically and electronically.  While the four valence electrons of \(2s^22p^2\) per carbon are distributed in the \(sp^2\) hybrid orbitals for the \(\sigma\) bond responsible for the triangular coordination in both 1D \textit{trans}-PA and 2D graphene, the remaining unpaired electron in \(2p_z\) is free to form a \(\pi\) bond between the nearest-neighbor carbon atoms for both, and sharing, depending on the coordination numbers, of two for 1D \textit{trans}-PA and three for 2D graphene.  If we substitute the H in 1D \textit{trans}-PA of (CH)\(_n\) with C and link the imaginary 1D ``(C\(_2)_n\)'' compounds along the interchain direction, it is clear that 1D \textit{trans}-PA (CH)\(_n\) is transformed into  2D graphene of (C\(_2\))\(_n\).\cite{Gutzler:2013ek}  In particular, the original conjugated \(\pi\)-bond system shared by two neighboring C-C pairs along the chain direction for 1D \textit{trans}-PA is now shared by three neighboring C-C pairs in a plane, as illustrated in Fig.~\ref{fig-PAtoGr}(c).  Considering the similarity in \(\pi\)-bond distribution due to limited paths of electron sharing,  1D \textit{trans}-PA and 2D graphene may also share the identical conducting mechanism due to the existence of similar conjugated \(\pi\)-bond systems, where the term of ``fractional charge'' in the 1D and 2D conduction bands could be interpreted as the tunneling of massless Dirac fermions across the Dirac point or \(\pi\)-bond sharing.\cite{Fasolino:2007bz, Kivelson:2001js}

\section{Graphene: two-dimensional topological material}

\subsection{Valence-bond model of graphene}

\begin{figure}
\begin{center}
\includegraphics[width=4.5in]{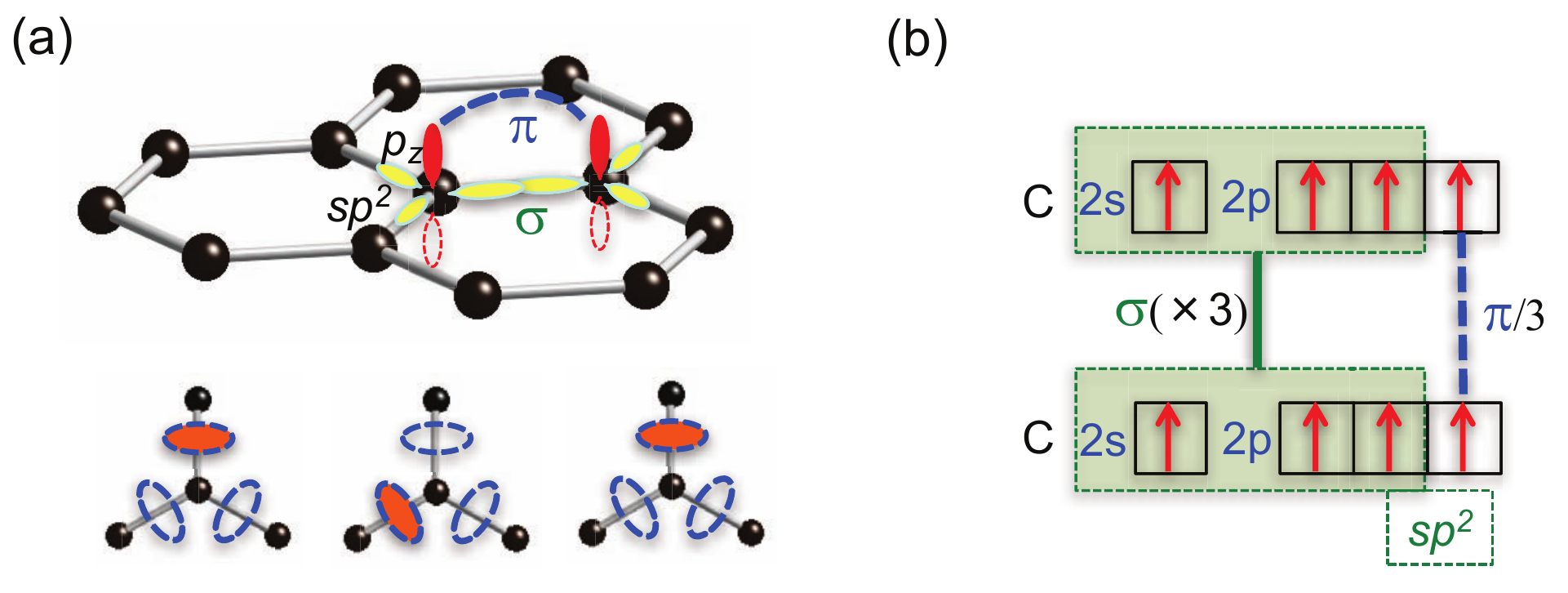}
\end{center}
\caption{\label{fig-graphene} (a) Crystal structure of 2D graphene, where the hybridized \(sp^2\) orbital per carbon is constructed into a honeycomb lattice of \(\text{CN}=3\) via \(\sigma\)-bond formation, and the remaining valence electron in \(2p_z\) forms a \(\pi\) bond with one of the three neighboring carbons to form a conjugated \(\pi\)-bond system.   }
\end{figure}

Following the same valence-bond analysis for 1D \textit{trans}-PA shown in Fig.~\ref{fig-PA}, the valence-bond model for graphene is illustrated in Fig.~\ref{fig-graphene}.  The valence-shell electrons in \(2s^22p^2\) hybridize into an \(sp^2\) orbital for the \(\sigma\)-bond formation among three neighboring carbon atoms in the honeycomb network, and the remaining unpaired electron in \(2p_z\) per carbon is proposed to form a side-to-side orbital overlap as a \(\pi\) bond.  Since the \(\pi\) bond must be shared by three equivalent neighboring carbon pairs in addition to the skeleton of a honeycomb lattice built by the three covalent \(\sigma\) bonds, distortion from the original triangular symmetry due to condensation is expected when the conjugated \(\sigma+\pi\) double bond exists randomly among one of three C-C pairs at any instance.\cite{Novoselov:2011gq} The broken local triangular symmetry could be absorbed into a 2D triangular supersymmetry when the distortion-restoration of the \(\pi\)-bond condensate fluctuates in time as a result of   quantum or thermal fluctuations.  

\begin{figure}
\begin{center}
\includegraphics[width=4.5in]{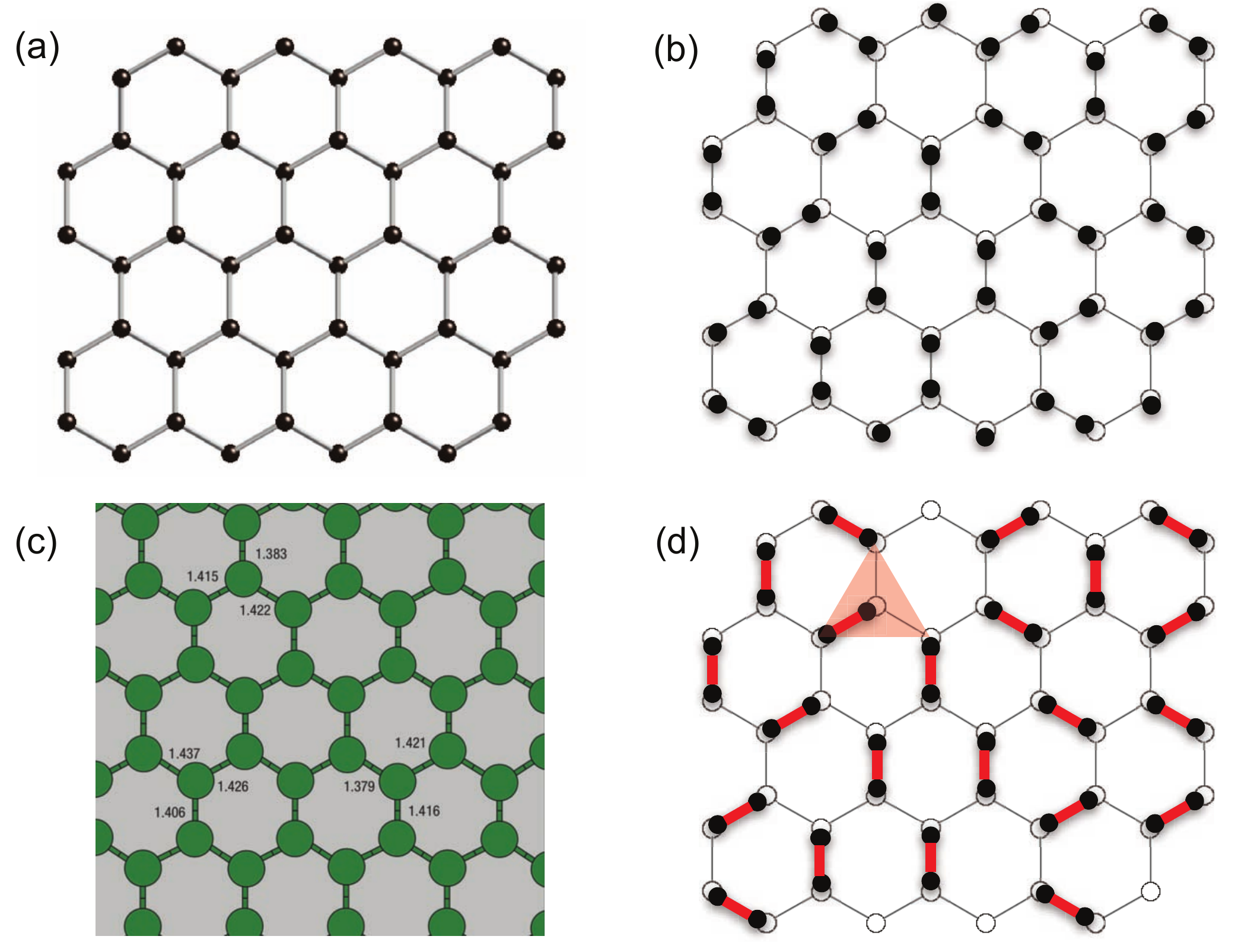}
\end{center}
\caption{\label{fig-eCrystal} (a) When only \(\sigma\)-bond electron clouds are considered among all C-C pairs in graphene, the 2D electronic crystal of homogeneous condensation has a symmetry identical to that of a 2D atomic crystal. (b) One of the infinite replicas of \(\sigma+\pi\) double-bond-induced inhomogeneous condensation to break the original symmetry. (c) Atomistic Monte Carlo simulation of bond length due to multiplicity of chemical bonding among carbon in graphene (reproduced from Ref. [\onlinecite{Fasolino:2007bz}] with permission; copyright © 2007, Springer Nature). (d) One of the infinite configurations of \(\pi\)-bond electrons shown by the red line, and the conjugated system requires a  \(C_3\) supersymmetry to absorb the constant distortion-restoration of an entangled network under quantum fluctuation.  }
\end{figure}

To explain the origin of the Dirac cone from the valence-bond perspective, we should first distinguish the atomic crystal and electronic crystal for graphene, as compared in Fig.~\ref{fig-eCrystal}.  Consider the valence electrons distributed near the C-C center: the electron cloud of \(\sigma\)-bond electrons form a 2D electronic crystal with a symmetry identical to that of the undistorted atomic crystal, as seen by the crystals formed with atoms (circles) and bonds (bars) of \(30^\circ\) rotation around the honeycomb center, as shown in Fig.~\ref{fig-eCrystal}(a).  Conversely, the conjugate nature of  the \(\pi\) bond with concomitant Peierls condensation would break the crystal symmetry under thermal or quantum fluctuation, as illustrated in Fig.~\ref{fig-eCrystal}(b) for one of the infinite degenerate and entangled configurations.  In fact, ripples of suspended graphene have been shown to be a common phenomenon as a result of the mixture of \(\sigma\) and \(\sigma+\pi\) bonding of different degrees of condensation, and the strain induced by the bond-length difference has also been demonstrated by an atomistic Monte Carlo simulation,\cite{Fasolino:2007bz} as seen in Fig.~\ref{fig-eCrystal}(c).  However, the slight distortion-restoration could be absorbed by the choice of a new basis highlighted by the red triangle of \(C_3\) symmetry [see Fig.~\ref{fig-eCrystal}(d)], instead of the static basis of 2C without distortion in the honeycomb network of the rhombic (hexagonal) Bravais lattice. In addition, the \(C_3\) symmetry can be viewed as a supersymmetry to protect the degenerate configurations in real space, which must be reflected in the band picture as a topological nontrivial state having symmetry-protected topological orders. The \(C_3\) crystalline-symmetry--protected topological orders in real space are similar to the time-reversal symmetry that protects the topological orders of the 1D MG chain in real space [see Figs.~\ref{fig-chains}(b) and \ref{fig-chains}(c)].

\subsection{Two-dimensional electron crystal and Dirac cone of graphene}

\begin{figure}
\begin{center}
\includegraphics[width=4.5in]{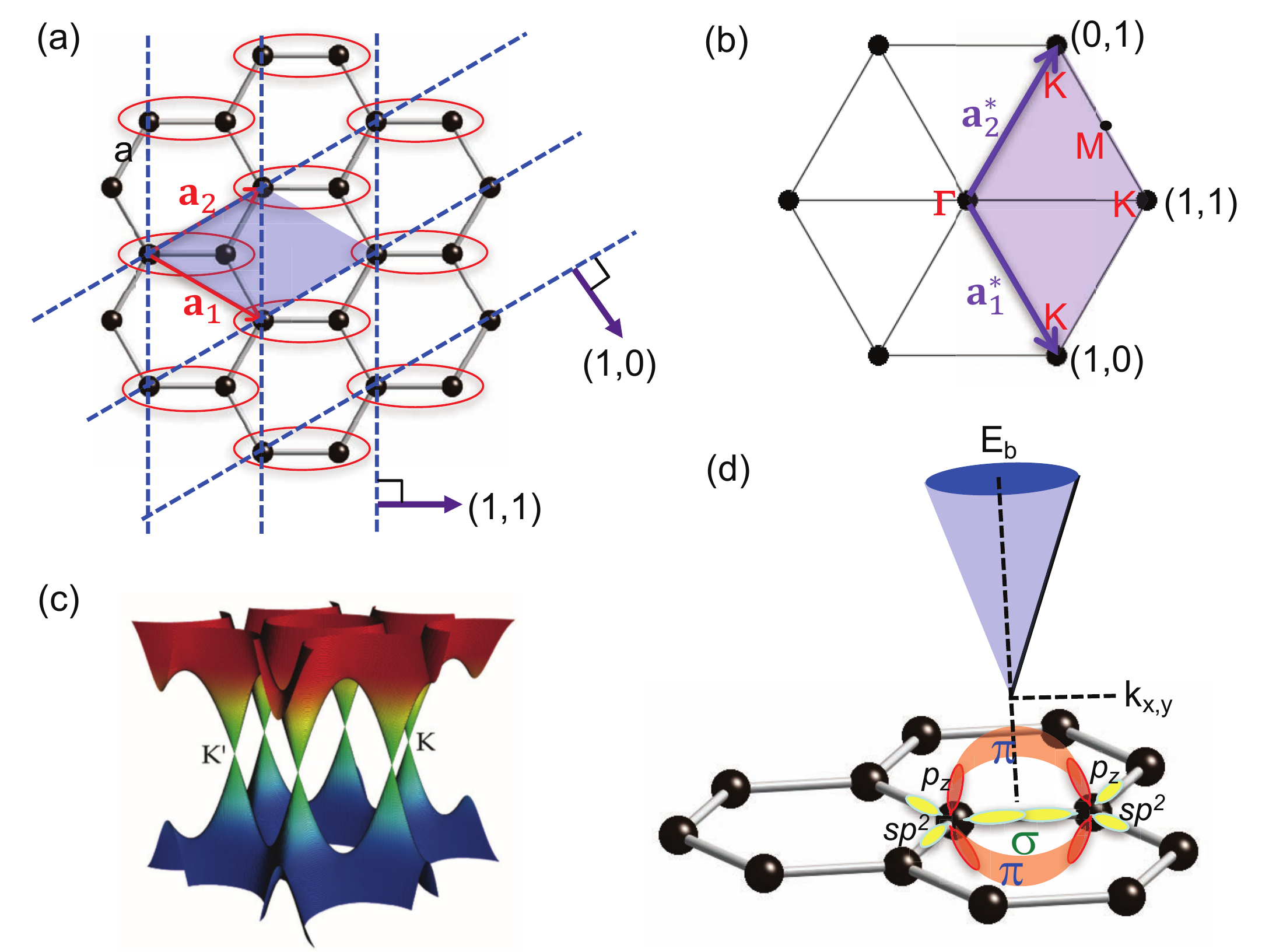}
\end{center}
\caption{\label{fig-DiracCone} (a) The crystal structure of graphene in a honeycomb lattice should be described with a 2D Bravais lattice of rhombic (hexagonal) unit cell with 2C (in red circle) as the basis. (b) The reciprocal lattice of graphene is identical to the real-space lattice with symmetry preserved by the Fourier transform in 2D, where each \(k\) point corresponds to the normal of lines of 2C bases in real space with specified \(d_{hk}\) spacing.  (c) The  graphene band picture are commonly taken out from a 2D cut of bulk band picture of graphite near the Dirac points labeled with two sets of \(K\) and \(K^\prime\) points. (d) The Dirac cone is interpreted as the \(E(k)\) linear dispersion for the binding energy of \(\pi\)-bond electrons from sitting  near the bond center (lowest) to near the atom center (highest). [Panel (c) is reproduced from Ref. [\onlinecite{Katsnelson:2007tu}] with permission; copyright © 2007 Elsevier Ltd.] }
\end{figure}

For graphene having a honeycomb lattice, its reciprocal lattice is identical to the real-space lattice of (rhombic) hexagonal symmetry, except that each real-space lattice point represents a 2C basis, so there is no need to distinguish the 2C basis into A and B carbons leading to two separate \(K\) and \(K^\prime\) points.\cite{Wallace:1947dy}   
The reason that the honeycomb lattice is not a 2D Bravais lattice is that the direct choice of unit vectors \(\textbf{a}_1\) and \(\textbf{a}_2\) for the two carbon atoms oriented at \(120^\circ\) would not generate all atomic positions via an integer number of translations along the unit axes [see Fig.~\ref{fig-DiracCone}(a)].  Crystallographically, a honeycomb lattice should be described with a Bravais lattice having a proper choice of basis; i.e., a \(60^\circ\) or \(120^\circ\) rhombic (hexagonal) Bravais lattice plus 2C as the basis for graphene, as shown in Fig.~\ref{fig-DiracCone}(a).  Since the reciprocal lattice must preserve its real-space symmetry via Fourier transformation, it is expected that the reciprocal lattice has an identical hexagonal unit cell.  Considering a 2D electron crystal formed only with \(\sigma\)- and \(\pi\)-bond electrons, a dynamic form factor for the 2C basis may tune the diffraction-peak intensity at each \textbf{k} point of the reciprocal lattice of hexagonal symmetry.  Although the confusing relationship between \(K\) and \(K^\prime\) points has been resolved by using a term named ``pseudospin'' to take care the bonding-nonbonding degeneracy,\cite{Fuhrer:2010dk} we believe that a detailed valence-bond analysis for graphene is still needed for a more intuitive understanding, especially regarding the origin of the Dirac cone.

Based on the 2D symmetry of graphene in real and reciprocal lattices, the 2D constructive diffraction spot shown at the \(K\) point is coming from the integrated intensity of \(\sigma\)- and \(\pi\)-bond electrons in lines, as illustrated in Fig.~\ref{fig-DiracCone}(a).  The Dirac-cone-shaped \(E(k)\) dispersion relationship for graphene at the \(K\) point can be mapped with the tunable incident beam energy.  Although both \(\sigma\) and \(\pi\) bonds contribute to the 2D constructive diffraction at the \(K\) point, it is expected that the \(\sigma\)-bond electrons have a larger binding energy at the level of a few eV  deeper in the valence band. The significantly lower binding-energy dispersion of the Dirac-cone shape near \(
{\sim}0.3\)--0.5 eV mapped by ARPES strongly suggests a massless tunneling phenomenon across the Dirac point is the  description mainly of \(\pi\)-bond electrons.\cite{Katsnelson:2007tu}  The conjugated nature of the \(\pi\) bond is due to the sharing among three neighboring C-C pairs under quantum fluctuation, which are responsible for the tunneling (bonding-nonbonding) behavior of \(\pi\)-bond electrons with an instantaneous symmetry breaking due to the bond-length change following condensation. The supersymmetry of \(C_3\) can absorb such a distortion when the original 2C basis is extended to 4C, as highlighted by the red triangle shown in Fig.~\ref{fig-eCrystal}(d).  

Although the electrons of positive binding energy sitting below the Fermi level means localization without metallic conduction following the Fermi-liquid theory, it does not apply to the 2D conduction mechanism for graphene.  The nonbonding state of \(\pi\) electrons are not itinerant electrons acting as the charge carriers for metallic conduction.  We find that graphene has a confirmed 2D conduction mechanism with a Dirac point sitting below the Fermi level, which suggests that the 2D conduction does not rely on the migration of itinerant electrons but rather on energy exchange through entangled bonding-nonbonding electrons of the conjugated \(\pi\)-bond system.  In particular, we should consider the particle-wave duality of electrons for the interpretation of the pairing and tunneling phenomenon via wave-function overlap.  Based on the argument above, we propose that the Dirac-cone-shaped band for graphene could be a description of the \(E(k)\) dispersion relationship for the electrons in \(\pi\) bonds distributed above and below the 2D basal plane, as shown in Fig.~\ref{fig-DiracCone}(d).  For each C-C pair, the binding energy of \(\pi\)-bond electrons changes from near zero at the C-C midpoint to a maximum near the carbon center, as indicated by the proximity to the defined \(K\) point of the linear \(E\)-\(k\) dispersion (\(E\)~\(\propto\)~\(k\)) drawn in a cone.  In addition, there are two lobes per \(p_z\) orbital for the \(\pi\)-bond bonding-nonbonding fluctuation, which is also consistent with the massless tunneling across the Dirac point to form two inverted cones.  Most importantly, the existence of gapless Dirac cones simply suggests that all electrons are confined in a plane via \(\pi\) bonding-nonbonding with constant tunneling behavior under quantum fluctuation of the lowest Gibbs free energy, in contrast with electrons in the bulk (\(E\)~\(\propto\)~\(k^2\)) that are either localized in the valence band or free in the conduction band via \(\sigma\) bonding of lowest enthalpy.  

\section{Topological insulators}

The unusual low-dimensional conductivity found in 1D \textit{trans}-PA and 2D graphene has been attributed to the existence of a conjugated \(\pi\)-bond system having an energy band with a linear energy-momentum dispersion relationship in the shape of a Dirac cone.  The key component of this finding is the identification of a conjugated \(\pi\)-bond system via a valence-bond analysis.  In particular, the formation of a conjugated \(\pi\)-bond system requires special conditions that allow a supersymmetry-protected electron sharing and structure distortion-restoration due to Peierls condensation, including atomic size, number of valence-shell electrons, bond length, and symmetry of local coordination.  More 3D TIs showing both insulating properties for the bulk and conducting properties for the surface have since been proposed and identified.\cite{Hasan:2010kua}  While the key signature of a 3D TI is the gapless Dirac-cone-shaped 2D band besides the gapped bulk band, the question remains of whether  a direct link exists between the 3D TI and  low-dimensional materials, such as 1D \textit{trans}-PA and 2D graphene, with common signatures.  In particular, if similar conjugated \(\pi\)-bond systems can also be identified in the known 3D TIs via valence-bond analysis in real space, it would provide the missing link to justify why and how a conjugated \(\pi\)-bond system is responsible for the surface-conduction mechanism of 3D TIs.  In the following, we discuss four representative classes of materials in the developmental history of TIs, including the CdTe-HgTe-CdTe quantum well that was used to probe the quantum spin Hall effect at the edge state, the strain-driven TIs of \(\alpha\)-Sn and Bi\(_{1-x}\)Sb\(_x\), the \(Z_2\)-type TI Bi\(_2\)Se\(_3\), and the topological crystalline insulator Pb\(_{1-x}\)Sn\(_x\)Se.

\subsection{Hg\(_{1-x}\)Cd\(_x\)Te: Tensile-strain-driven topological insulator state}

\begin{figure}
\begin{center}
\includegraphics[width=5.5in]{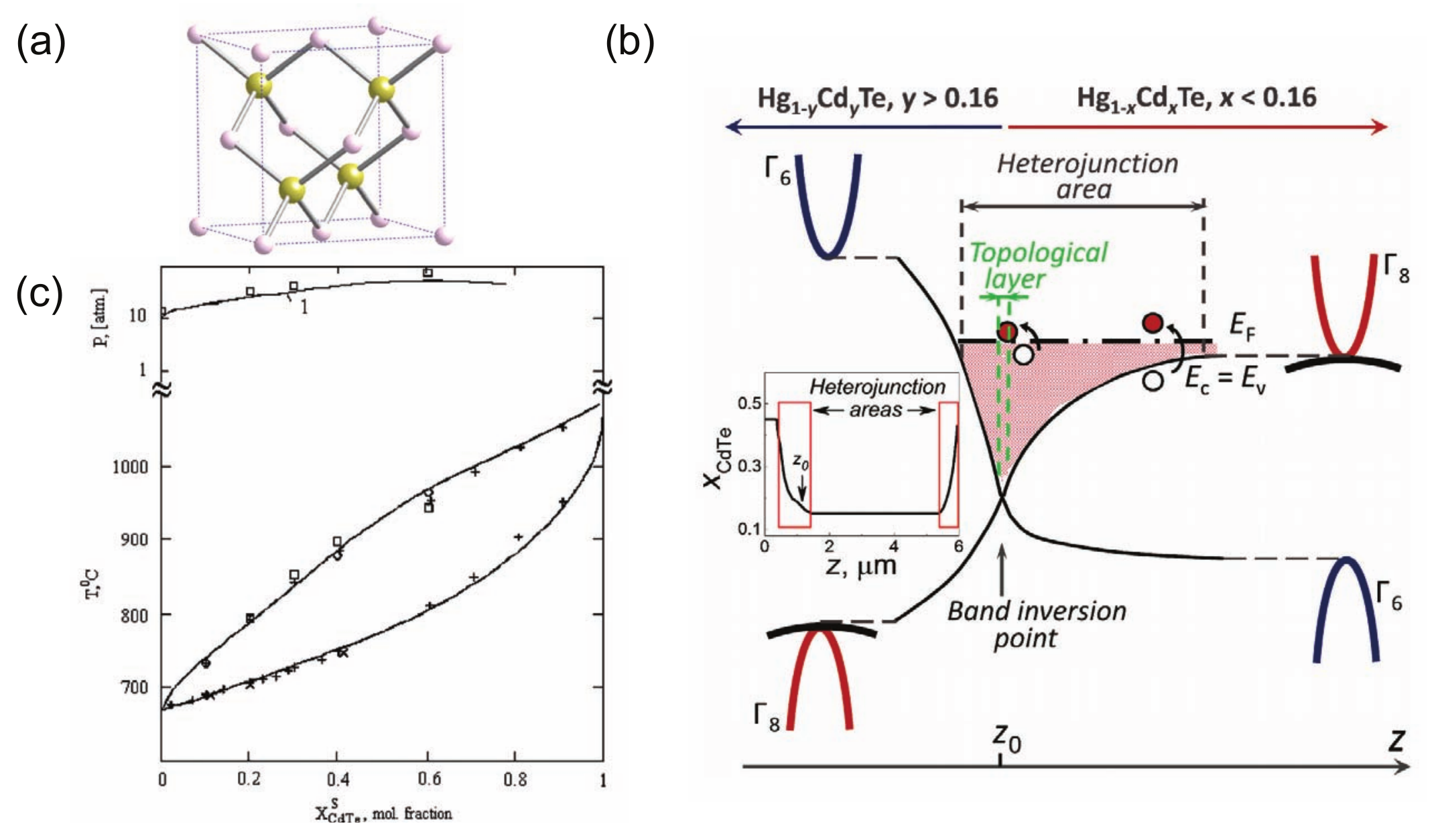}
\end{center}
\caption{\label{fig-HgCdTe} (a) The crystal structure of Hg\(_{1-x}\)Cd\(_x\)Te is a zinc-blende type fcc structure of space group \(F\bar{4}3m\), with Hg at (\(0,\,0,\,0\)) and Te at (\(\frac{1}{4},\,\frac{1}{4},\,\frac{1}{4}\)). (b) Band inversion occurs near the HgTe-CdTe interface of the quantum well. (c) HgTe-CdTe binary phase diagram shows the full solubility, and the liquid + solid mixture between the liquidus and solidus indicates that  a strain energy is stored in the system.  [Panel  (b) is reproduced from Ref. [\onlinecite{Galeeva:2018fx}], and panel (c) is from Ref. [\onlinecite{Moskvin:2008fo}] with permission; Copyright 2018, Elsevier.]  }
\end{figure}

The quantum spin Hall effect has been proposed to distinguish  TIs from  conventional insulators for their dissipationless spin current carried by  helical edge states, as was theoretically proposed and later verified with the CdTe-HgTe-CdTe quantum-well experiment.\cite{Bernevig:2006ij, Konig:2007hs}  The normal band gap between the \(s\)-type \(\Gamma_6\) and the \(p\)-type \(\Gamma_8\) for semiconducting CdTe is inverted at \(x_c \sim0.16\) for Hg\(_{1-x}\)Cd\(_{x}\)Te toward semimetal HgTe, and the quantum spin Hall effect is confirmed by the measured contact conductance of \(2e^2/h\) for a quantum well with thickness only larger than \(
{\sim}64\)~\AA, i.e., showing surface electron states without being smeared by the bulk state of the  CdTe-CdTe wave function tunneling across the well at the interface near \(x_c \sim0.16\), as shown in Fig.~\ref{fig-HgCdTe}(b).  The main reason for this experimental design is for its tunable band of Hg\(_{1-x}\)Cd\(_x\)Te solid solution with band inversion at a critical level and the heavy elements of inherited strong SOC.  For the purpose of valence-bond analysis, it is instructive to examine the evolution of crystal symmetry, lattice size, and melting point as a function of \(x\), as shown in Fig.~\ref{fig-HgCdTe}(b). Because both HgTe and CdTe have the identical zinc blende type fcc crystal structure with Hg or Cd at (\(0,\,0,\,0\)) and Te at (\(\frac{1}{4},\,\frac{1}{4},\,\frac{1}{4}\)),  both Hg and Cd are in  group 12 with similar electronic configuration so that the solubility of Hg\(_{1-x}\)Cd\(_x\)Te is high from \(x=0\) to 1 without undergoing a structural phase transition.\cite{Rogalski:2005ju}  It is important to note that the semimetal HgTe has an indirect band overlap (negative gap of \(
{\sim} -0.26\) eV) and CdTe is a semiconductor of larger gap of \(
{\sim}1.49\) eV, which is also consistently reflected in the low melting point of \(
{\sim}680\,^\circ\)C for HgTe and the much higher melting point of \(
{\sim}1090\,^\circ\)C for CdTe.  In particular, band inversion occurs smoothly near the critical level of \(x_c \sim0.16\) to transform the system from a semimetal with a single Fermi node to a system with a narrow band gap of strong SOC, which is favorable for the formation of a topological state.  We might consider the SOC mechanism across the narrow gap to be analogous to a two-level system under thermal fluctuation; i.e., the electrons at the top of the valence band  exchange energy between the triplet Zeeman energy gain (loss) and the electron potential loss (gain), as if the edge state (2D plane for the 3D system) is threaded via chiral electrons (spins). For the TI having a topologically nontrivial state, a hidden supersymmetry must be present to absorb the impact of distortion-restoration of the bond lengths that instantaneously break the local crystal symmetry. 

\begin{figure}
\begin{center}
\includegraphics[width=5.5in]{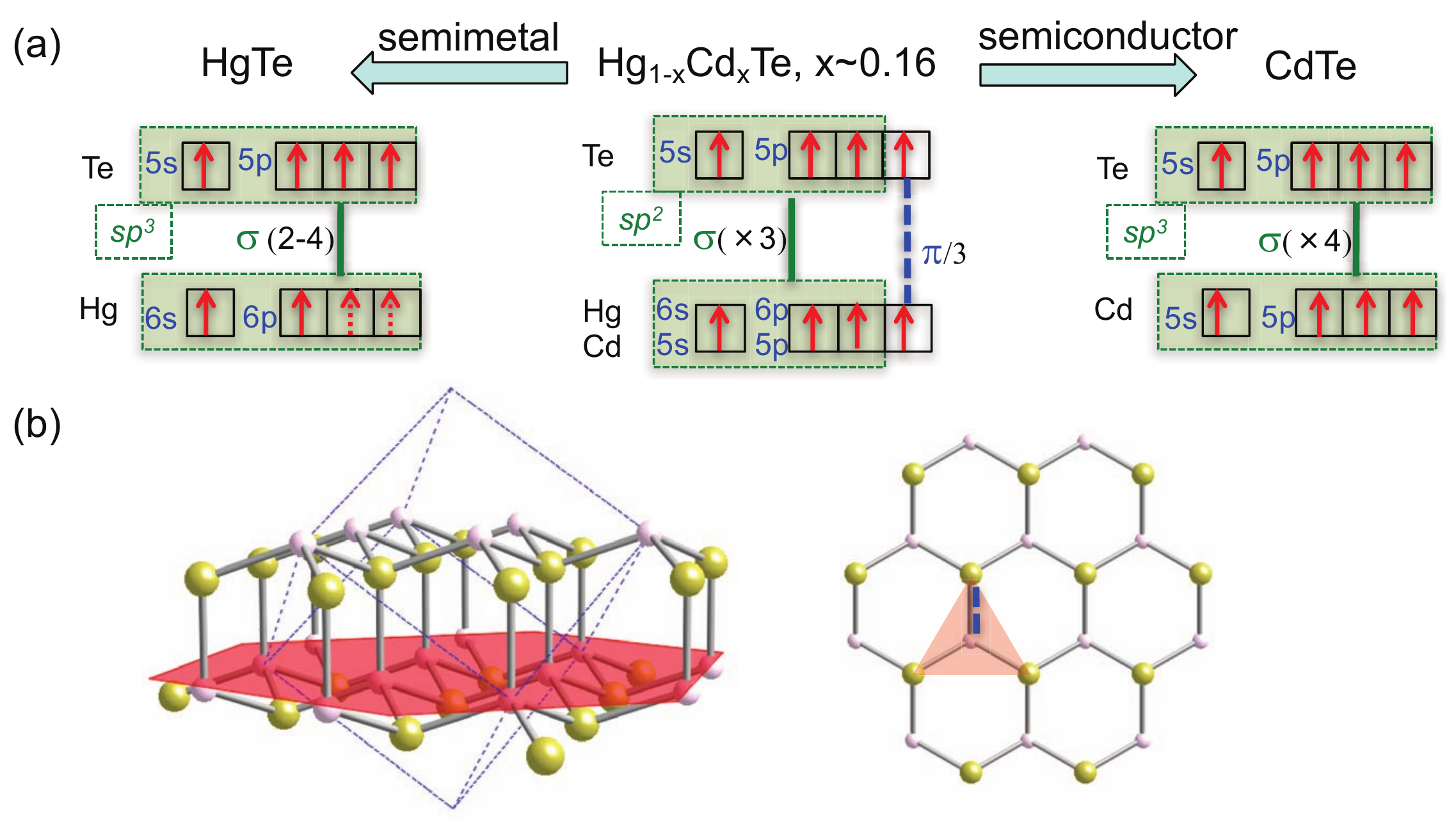}
\end{center}
\caption{\label{fig-HgTe111} (a) The evolution of electronic configuration from HgTe to CdTe with enhanced tensile strain. (b) For Hg\(_{1-x}\)Cd\(_x\)Te of x\(\gtrsim\)0.16 with a narrow band gap, 2D states of conjugated \(\pi\)-bond system via \(sp^2\) orbital hybridization becomes possible.}
\end{figure}

To explore why and how band inversion occurs near \(x_c \sim0.16\) for the solid solution of Hg\(_{1-x}\)Cd\(_x\)Te, several valence-bond models of HgTe-CdTe are proposed, as shown in Fig.~\ref{fig-HgTe111}(a).  Based on the atomic electron configurations of Hg([Xe]\(4f^{14}5d^{10}6s^2\)), Cd([Kr]\(4d^{10}5s^2\)), and Te([Kr]\(4d^{10}5s^25p^4\)), the gapped CdTe of fcc structure with a higher melting point is proposed to have stronger covalent \(\sigma\) bonds among the tetrahedrally coordinated Cd-Te via hybridized \(sp^3\) orbitals (with two electrons donated from Te to Cd).  HgTe is categorized as a semimetal because of indirect band overlap,\cite{Rogalski:2005ju} which corresponds to the two electrons donated from Te being partially localized in the \(sp^3\) hybridized orbital of Hg.  The difference in bonding character  between HgTe and CdTe is also closely related to the same principal quantum number of \(n=5\) for Cd and Te for more congruency, in contrast with Hg and Te, which have different \(n\).  Whereas the lattice constant increases with \(x\) for a more rigid \(\sigma\)-bond formation, increasing tensile-strain energy is expected for higher \(x\) until reaching the critical level of Cd substitution at \(x_c \sim0.16\), which could be used to initiate the SOC mechanism for samples with \(x\gtrsim0.16\) having narrow gaps to become  TIs.  

In the topologically nontrivial state,  an electronic dimensional crossover is proposed to be responsible for the 2D conduction mechanism as a result of the existence of the  conjugated \(\pi\)-bond system, which requires that the electronic configuration shift statistically from the 3D \(sp^3\) hybridization to the 2D \(sp^2\) hybridization along the (111) direction, so that the unpaired electron per \(p_z\) of both Hg or Cd and Te may form a \(\pi\) bond via side-to-side orbital overlap of neighboring half-filled \(p_z\) orbitals. However, each \(\pi\) bond must be shared statistically by three neighboring identical Hg(Cd)-Te pairs  as a conjugated system [Fig.~\ref{fig-HgTe111}(b)].  Because of the coexisting Peierls condensation for \(\pi\)-bond formation of paired electrons in singlet state, the induced crystal-symmetry breaking in the form of local distortion-restoration in time can be absorbed by a \(C_3\) supersymmetry.  Comparing the similar electronic configurations of (111) projection of corrugated Hg\(_{1-x}\)Cd\(_x\)Te at the \(x\sim0.16\) plane with that of graphene (Fig.~\ref{fig-eCrystal}), it is not surprising that Kane and Mele earlier identified the quantum spin Hall effect in graphene.\cite{Kane:2005hl}

\subsection{\(\alpha\)-Sn: Compressive-strain-driven topological insulator state}

\begin{figure}
\begin{center}
\includegraphics[width=5.5in]{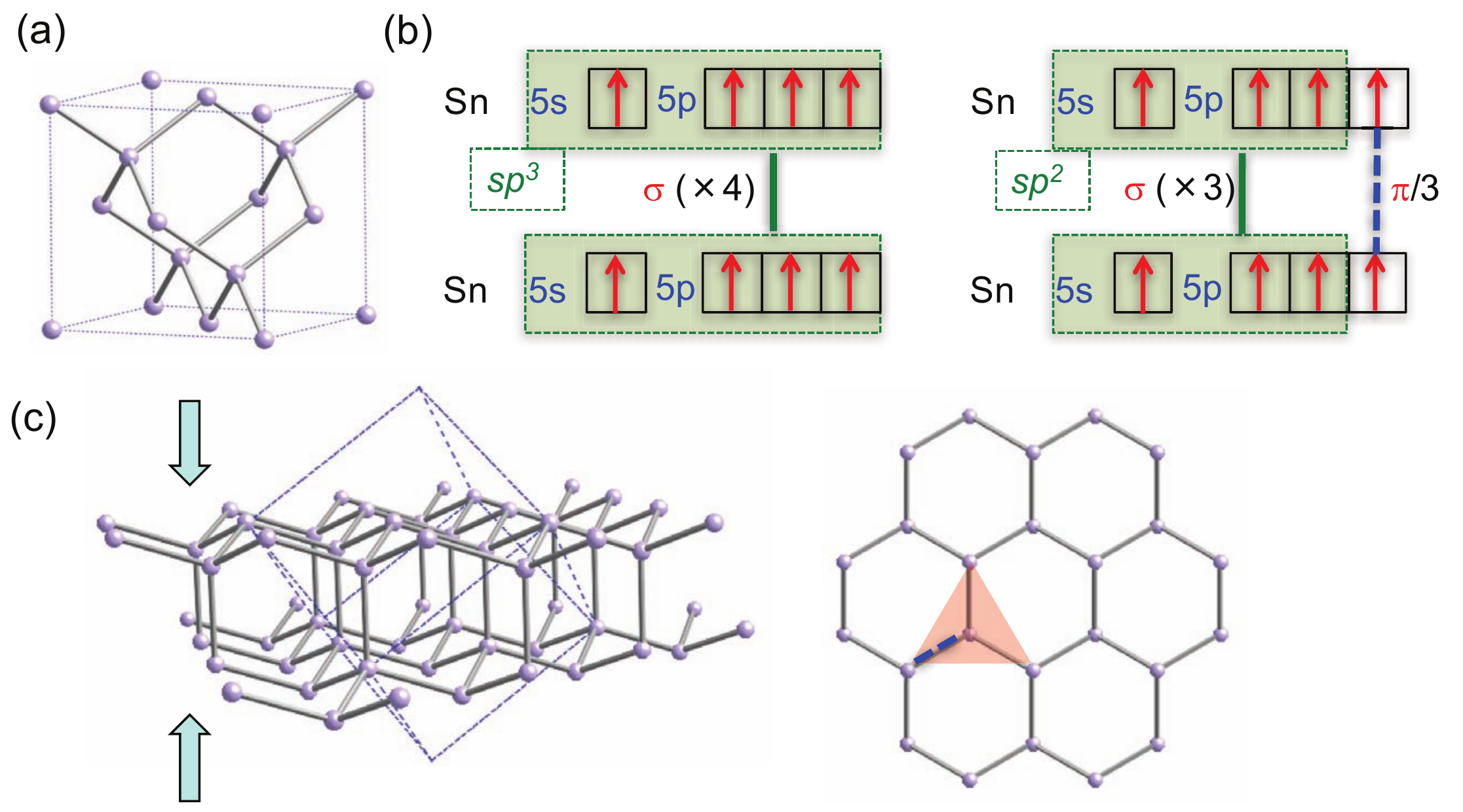}
\end{center}
\caption{\label{fig-alphaSn} (a) The crystal structure of \(\alpha\)-Sn is similar to that of diamond with the space group \(Fd\bar{3}m\). (b) The valence-bond models using valence electrons distributed in \(sp^3\) hybrid and \(sp^2\) hybrid plus a conjugated \(\pi\)-bond system. (c) Under  uniaxial compressive strain, 2D-like electronic crossover is preferred with the help of added quasi-2D \(\pi\) bonds. }
\end{figure}

\(\alpha\)-Sn has a crystal structure similar to that of diamond and Si in a fcc-type of space group \(Fd\bar{3}m\) (No. 227).  The cubic degeneracy of \(\alpha\)-Sn can be removed under uniaxial compression along the (111) direction to exhibit band anisotropy,\cite{Roman:1972fc} and this material has also been predicted to be a strong TI under uniaxial strain.\cite{Fu:2007ei}  We find that \(\alpha\)-Sn is a perfect example to demonstrate the generation of a topological state via electronic dimensional crossover.  The atomic electron configuration of Sn ([Kr]\(4d^{10}5s^25p^2\)) and its tetrahedral coordination suggests that the four valence-shell electrons are hybridized into \(sp^3\) to form four weak \(\sigma\) bonds with neighboring Sn atoms, as reflected by its narrow band gap. This is also consistent with the low-temperature phase of semiconducting \(\alpha\)-Sn (gray tin) to the high-temperature metallic phase of \(\beta\)-Sn (white tin) across the structural transition temperature near \(
{\sim}13\,^\circ\)C.\cite{Bassani:1963kr}   The valence-bond models of \(\alpha\)-Sn with and without uniaxial strain are compared in Fig.~\ref{fig-alphaSn}(b), where the original narrow-gap semiconducting state has a covalent \(\sigma\)-bond system of tetrahedral shape with the help of \(sp^3\) hybridization, so the weak covalent-bond nature must come from its high principal quantum number of \(n=5\) with low binding energy.  As a uniaxial pressure is applied along the (111) direction, the system may be viewed as a quasi-2D system with weakened interlayer coupling as a result of the added Coulomb repulsion among layers, which leads to a reasonable assumption that \(\sigma\) bonds with \(sp^2\) hybridization in the (111) plane are favored statistically, as shown in Fig.~\ref{fig-alphaSn}(c).  

Interestingly, the remaining unpaired electron in \(5p_z\) may still  form a \(\pi\) bond as a result of concomitant Peierls condensation, which breaks the crystal symmetry, but the  three identical Sn-Sn pairs in the quasi-2D layer with electrons undergoing quantum fluctuation must force the \(\pi\) bond to be shared statistically as a conjugated system.  The current description of \(\alpha\)-Sn under compressive strain as a quasi-2D system is nearly identical to that of  perfect 2D graphene (Fig.~\ref{fig-eCrystal}), except that the plane is corrugated for the former.   In fact, we may extend the same valence-bond analysis to the isolation of quasi-2D compounds from the typical semiconductors Si and Ge of similar electron configuration and \(sp^3\) tetrahedral coordination via a uniaxial compression or mechanical exfoliation, i.e., the parallel compounds of graphene called silicene and germanene are proposed,\cite{Butler:2013ha} respectively. However, it is  more difficult to generate the 2D compounds of silicene and germanene due to their much stronger \(\sigma\) bonds, which lead to the larger band gap and higher melting point.  To obtain the topological state of a covalent compound with cubic structure by applying a uniaxial compression along the (111) direction for the close-packing planes, one must choose a weak \(\sigma\)-bond system or  create an internal strain via intercalation or layer mismatch.

\subsection{Bi\(_{1-x}\)Sb\(_x\): Compressive-strain-driven topological insulator state}

\begin{figure}
\begin{center}
\includegraphics[width=6.5in]{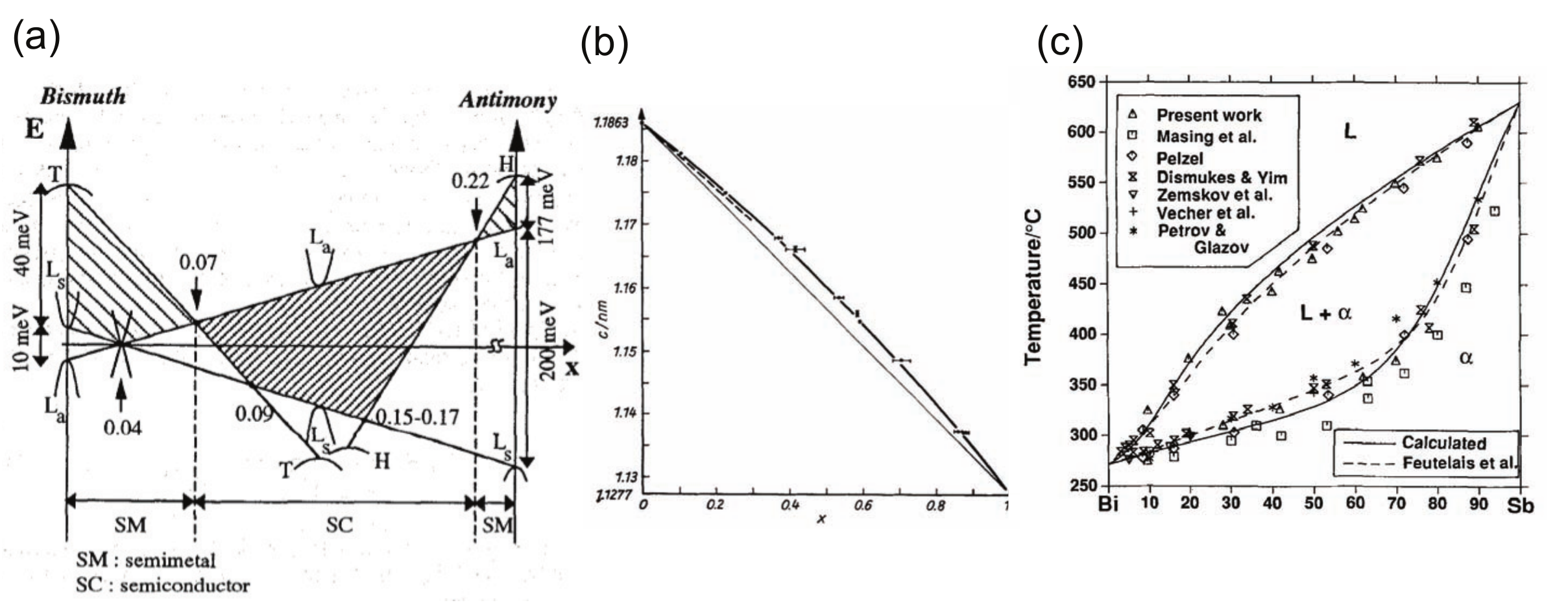}
\end{center}
\caption{\label{fig-BiSb} (a) The bulk band energy evolution of Bi\(_{1-x}\)Sb\(_x\) as a function of \(x\). Band inversion occurs near \(x\sim0.04\) and 0.07. (b) The lattice parameter decreases with increasing \(x\), but the \(c\) axis is shown deviating above  Vegard's prediction, which suggests weakened interlayer coupling. (c) Bi-Sb binary phase diagram indicates the full solubility and that Sb has a stronger \(\sigma\) bond, as reflected by the higher melting point.  The liquid + solid mixture suggests that strain energy is stored in the system.   (Reproduced from Refs. [\onlinecite{Lenoir:1998tl, Berger:1982ds, Ohtani:1994ix}] with permission; copyright © 1997, published by Elsevier, Ltd.;  copyright © 2006, John Wiley and Sons;  copyright © 1994, The Metallurgical of Society of AIME.) }
\end{figure}

\begin{table}
\caption{\label{table-BiSb} Bond length and bond angle of  Bi octahedron for Bi\(_{1-x}\)Sb\(_x\).\cite{Cucka:1962jr} Bond lengths \(l_i\) and bond angles \(\theta_i\) are defined in Fig.~\ref{fig-BiSbVB}(a).}
\begin{tabular}{cccc}
 \hline\hline
Sample & Bond lengths \(l_1\), \(l_2\) (\AA) & Bond angles \(\theta_1\), \(\theta_2\) (\(^\circ\)) & Category \\
 \hline
Bi & 3.064, 3.516 & 95.48, 80.31 & Semimetal\\
Bi\(_{0.8}\)Sb\(_{0.2}\) & 3.010, 3.563 & 97.23, 78.66 & Semiconductor\\
 
Bi\(_{1-x}\)Sb\(_{0.04-0.21}\) &  &  & Topological insulator\\
 \hline\hline
\end{tabular}
\end{table}

\begin{figure}
\begin{center}
\includegraphics[width=4.5in]{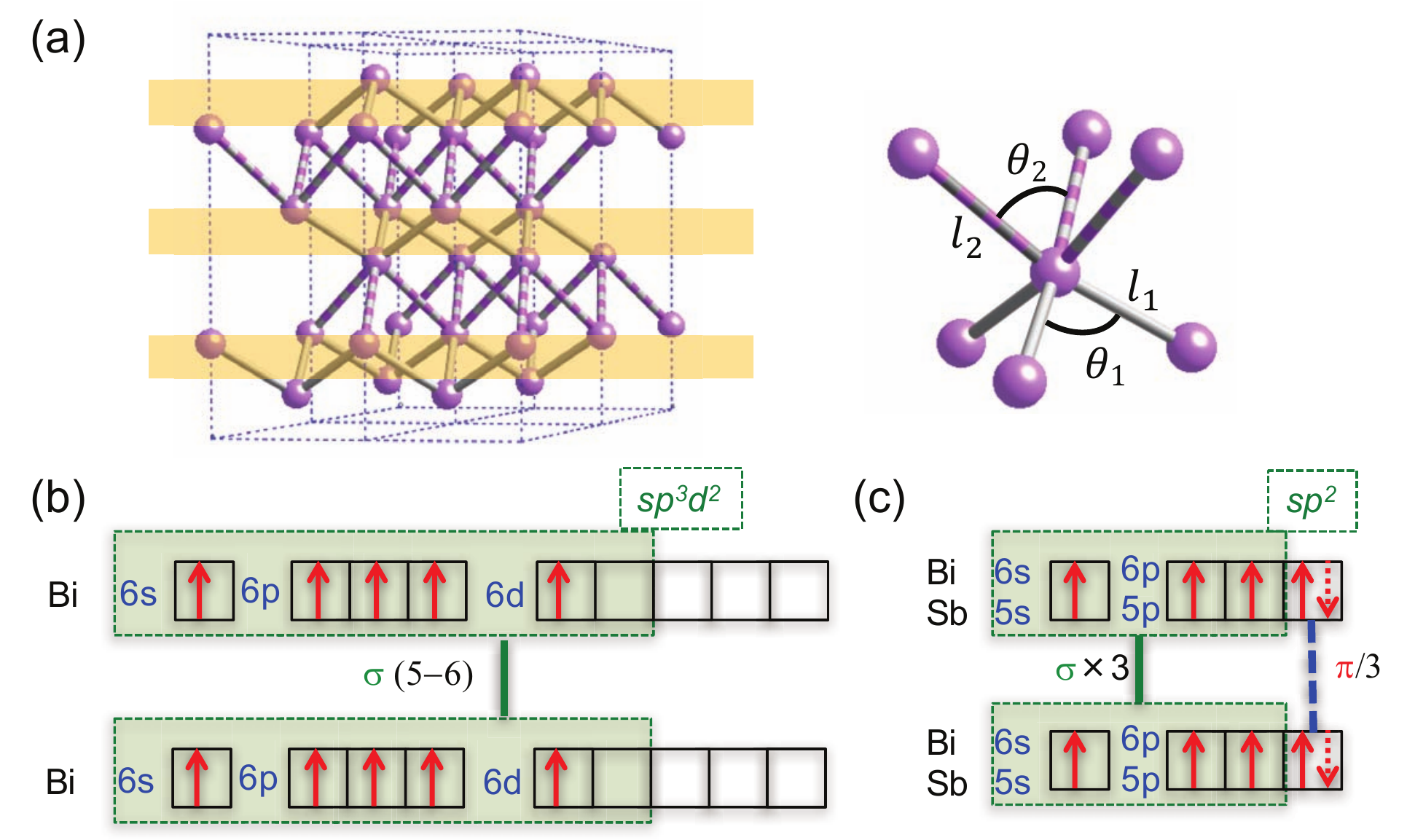}
\end{center}
\caption{\label{fig-BiSbVB} (a) The crystal structure of Bi\(_{1-x}\)Sb\(_x\) has a subtle 2D character via the slightly flattened lower half of each Bi octahedral coordinated layer.  See also Table \ref{table-BiSb}. (b) The valence-bond model for Bi crystal with covalent \(\sigma\) bonds via \(sp^3d^2\) hybridization. (c) The valence-bond model for Bi\(_{1-x}\)Sb\(_x\) (x\(\sim\)0.04-0.21) having a topological state, where covalent \(\sigma\) bonds are formed via \(sp^2\) hybridization with a conjugated \(\pi\)-bond system in quasi-2D. }
\end{figure}

Together with \(\alpha\)-Sn and HgTe under uniaxial strain, the Bi\(_{1-x}\)Sb\(_x\) alloy has also been predicted to be a  strong 3D TI,\cite{Fu:2007ei} as was   verified by ARPES experiments.\cite{Hsieh:2008tf, Nakamura:2011gb}  The Bi-Sb alloy is similar to that of HgTe-CdTe with full solubility and without an intermediary structural phase transition. Bi as a semimetal of indirect band overlap is converted to a semiconductor for \(x\sim0.07\)--0.22 with band inversion at the critical levels of \(x \sim0.04\) and 0.07, as shown in Fig.~\ref{fig-BiSb}(a).  Pure Bi has a rhombohedral crystal structure of space group \(R\bar{3}m\), which is a common space group to describe a layered compound of close packing per layer along the (111) direction.  However, the strong SOC for heavy Bi ([Xe]\(4f^{14}5d^{10}6s^26p^3\))  with a half-filled 6\(p\) orbital  creates a distortion within each layer via a Peierls-like dimerization, as seen by the flattened layer with the Bi octahedron of broken mirror symmetry (Fig.~\ref{fig-BiSbVB}), which could also contribute to the unusually large diamagnetism and spin Hall conductivity.\cite{Yuki:2014ek}  

The electronic structure of the Bi-Sb alloy is complex due to strong SOC among the narrowly split energy levels of weak binding energy for orbitals of high principal quantum number (\(n \gtrsim5\)), but a trend of dimensional crossover for the electronic structure from 3D to 2D is revealed by the evolution of bond length and bond angle with \(x\),\cite{Cucka:1962jr} as illustrated in Fig.~\ref{fig-BiSbVB}(a) and Table~\ref{table-BiSb}. The slight \(c\)-axis deviation that is greater than that predicted by Vegard's law also suggests a weakened interlayer coupling, as shown in Fig.~\ref{fig-BiSb}(b).  The comparison of bond length, bond angle, and lattice parameters between Bi and Bi\(_{0.8}\)Sb\(_{0.2}\) is summarized in Table~\ref{table-BiSb}, which suggests that Peierls condensation occurs for the lower-three \(\sigma\) bonds of the octahedral \(sp^3d^2\) hybrid orbital for pure Bi, increasing the level of smaller Sb ([Kr]\(4d^{10}5s^25p^3\)) substitution and creating stronger \(\sigma\) bonding (higher melting point) to provide additional internal compressive strain energy due to atomic-size mismatch, which must be stored in the system and used to initiate the SOC mechanism for samples of band inversion or narrow the gap in the range  \(x\sim0.04\)--0.21 as a TI.  

Because the Bi-Sb alloy is too complex, having various degrees of instantaneous local distortion that break the local crystal symmetry, we propose a valence-bond model for the two extreme cases of a 3D semimetal and a 2D semiconductor for the Bi-Sb alloy system, as shown in Fig.~\ref{fig-BiSbVB}(b)-(c).  It is important to note that  Bi and Sb have five electrons in the valence shell, so that the \(sp^3d^2\) hybridization in octahedral coordination must have one empty bond out of the six \(\sigma\) bonds [Fig.~\ref{fig-BiSbVB}(b)]; i.e., limited electron sharing (local charge transfer) is expected for the semimetal character for the bulk.  Following a similar argument of electronic dimensional crossover from 3D to 2D for the topological state of Hg\(_{1-x}\)Cd\(_x\)Te and \(\alpha\)-Sn under compressive strain, we expect that a similar conjugated \(\pi\)-bond system could also be identified in this \(x\) range, with the Dirac cone mapped by ARPES experiments.  Figure~\ref{fig-BiSbVB}(c) shows an ideal 2D configuration with a possible conjugated \(\pi\)-bond system for Bi\(_{1-x}\)Sb\(_x\) (x\(\sim\)0.04-0.21), where the valence-shell electrons are prone to form \(sp^2\) hybridization so that the remaining two electrons per \(p_z\) could either be filled or form a \(\pi\)-bond with the three neighboring  Bi or Sb atoms under quantum fluctuation, which leads to a configuration similar to that of graphene but  corrugated in quasi-2D.  The broken mirror symmetry within each layer [Fig.~\ref{fig-BiSbVB}(a)] is a reflection of the progressive electron dimensional crossover as a function of increasing Sb substitution of added internal compressive strain.

\subsection{Bi\(_2\)Se\(_3\): Time-reversal-symmetry--protected topological insulator state}

\begin{figure}
\begin{center}
\includegraphics[width=5.0in]{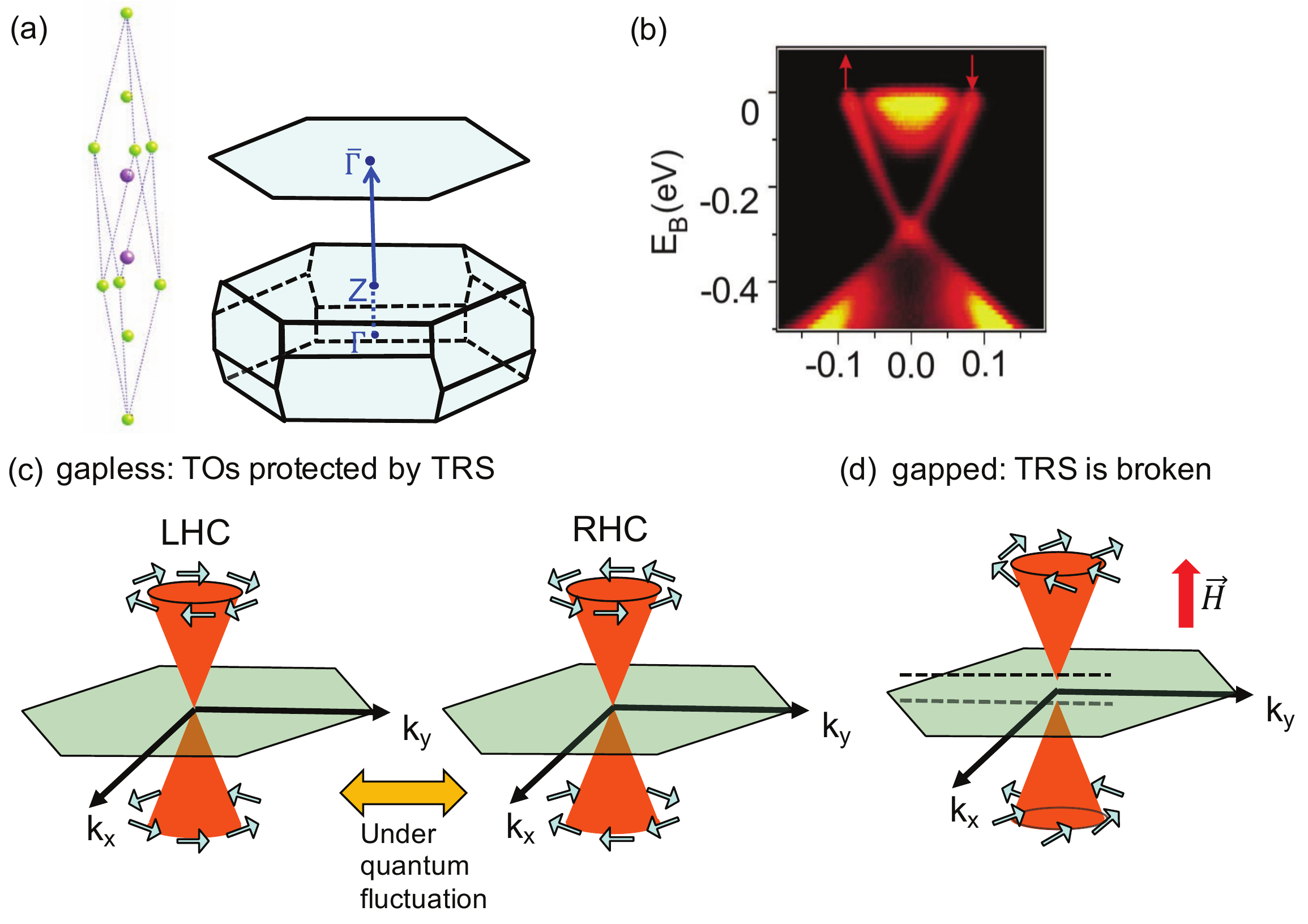}
\end{center}
\caption{\label{fig-chiral} (a) Bi\(_2\)Se\(_3\) can be described by the rhombohedral space group \(R\bar{3}m\), as seen by the rhombohedral primitive unit cell and its bulk Brillouin zone with surface projection in the \(\Gamma\)-\(Z\) direction. (b) The typical energy-momentum dispersion at the \(\Gamma\) point shows the gapped bulk band and the gapless spin-polarized surface band as a Dirac cone. (c) Topological surface states are protected by time-reversal symmetry (TRS), where spin chirality changes sign across the Dirac point.  (d) When the TRS is broken by a magnetic field, the Dirac cone becomes gapped for the nondegenerate ground state. [Panel (b) is reproduced from Ref. [\onlinecite{Hsieh:2009dp}] with permission; copyright © 2009, Springer Nature.] }
\end{figure}

Bi\(_2\)Se\(_3\) is the most widely researched TI and  is categorized as a \(Z_2\)-type TI with time-reversal-symmetry--protected (TRS--protected) topological orders with confirmed Dirac-cone existence along the \(\Gamma\)-\(Z\) direction, where quintuple layers composed of Se1-Bi-Se2-Bi-Se1 have van der Waals gaps in between, as described by the 3D crystal symmetry of space group \(R\bar{3}m\) in the hexagonal family,\cite{shu2018dynamic, Hasan:2010kua, Hsieh:2009dp} as shown in Fig.~\ref{fig-chiral}(a)-(b). The TRS-protected topological orders suggest that the spins in 2D are spin-momentum-locked helical spins of fixed chirality, as illustrated in Fig.~\ref{fig-chiral}(c), so that the TI state is robust against the impact of nonmagnetic-impurity inclusion.  However, once the TRS is broken by applying a local magnetic field from the ferromagnetic impurities on the surface of a Bi\(_2\)Se\(_3\) crystal, the Dirac-cone gap is open for the whole system to indicate the disruption of surface conduction under entanglement, as illustrated in Fig.~\ref{fig-chiral}(d).\cite{Hasan:2010kua}  Although the nature of the \(Z_2\)-type TI has been described in mathematical rigor by  first-principles calculations and a parity analysis\cite{Zhang:2009vu} and by a spin-resolved ARPES experiment that confirmed the chirality of the Dirac cone,\cite{Hsieh:2009dp} it is still too abstract without the supportive understanding of the topological character via a valence-bond analysis, especially with regards to the meaning of the TRS-protected topological orders.  

\begin{figure}
\begin{center}
\includegraphics[width=5.0in]{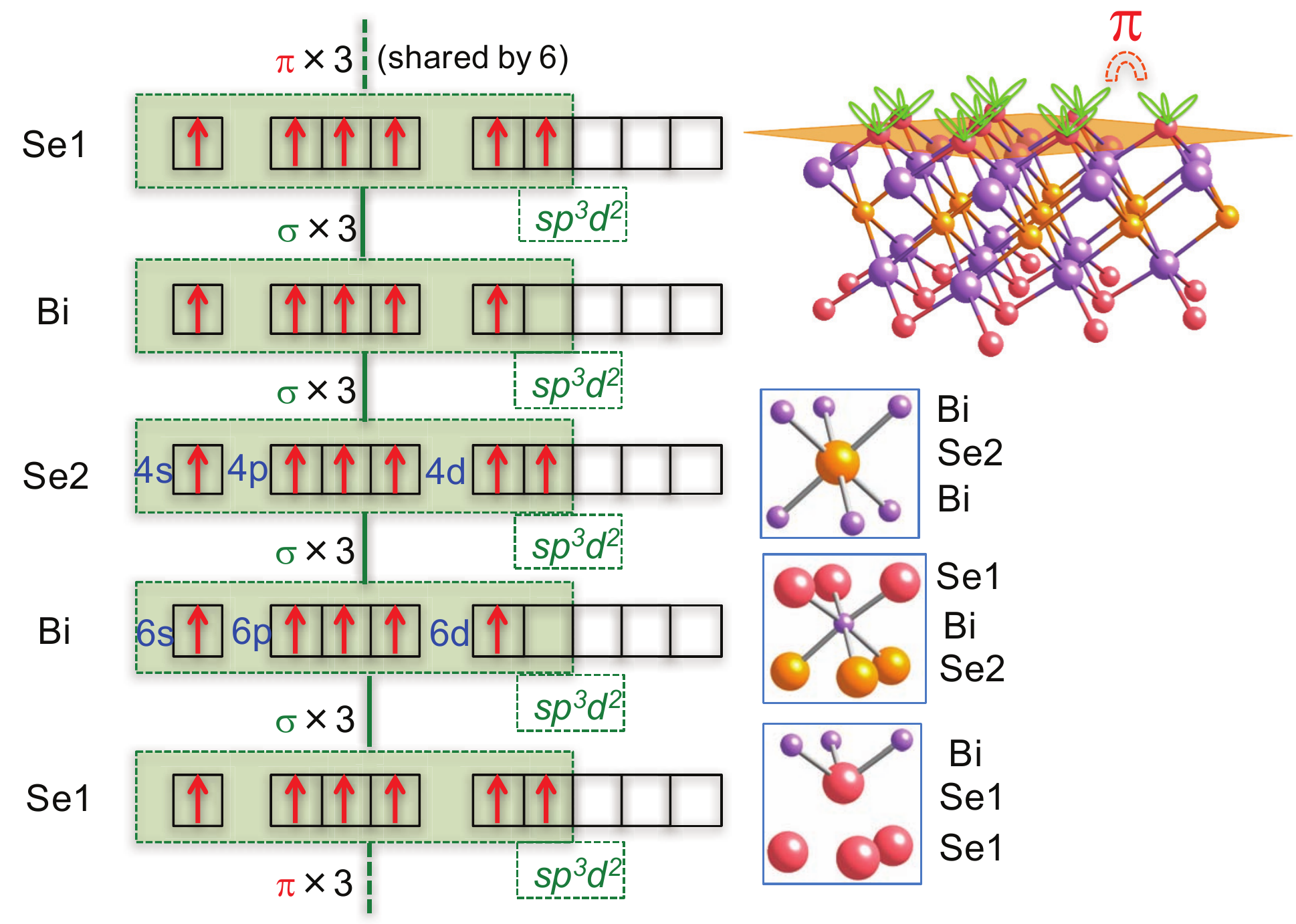}
\end{center}
\caption{\label{fig-Bi2Se3} The valence-bond model of Bi\(_2\)Se\(_3\), where all atoms in the quintuple layer are hybridized into \(sp^3d^2\) for octahedral coordination and form \(\sigma\) bonds in the bulk. The remaining three unpaired electrons per Se1 form three \(\pi\) bonds with the six neighboring Se1 atoms as a conjugated system. (Reproduced from Ref. [\onlinecite{shu2018dynamic}] with permission; copyright © 2018, APS.)}
\end{figure}

\begin{figure}
\begin{center}
\includegraphics[width=5.0in]{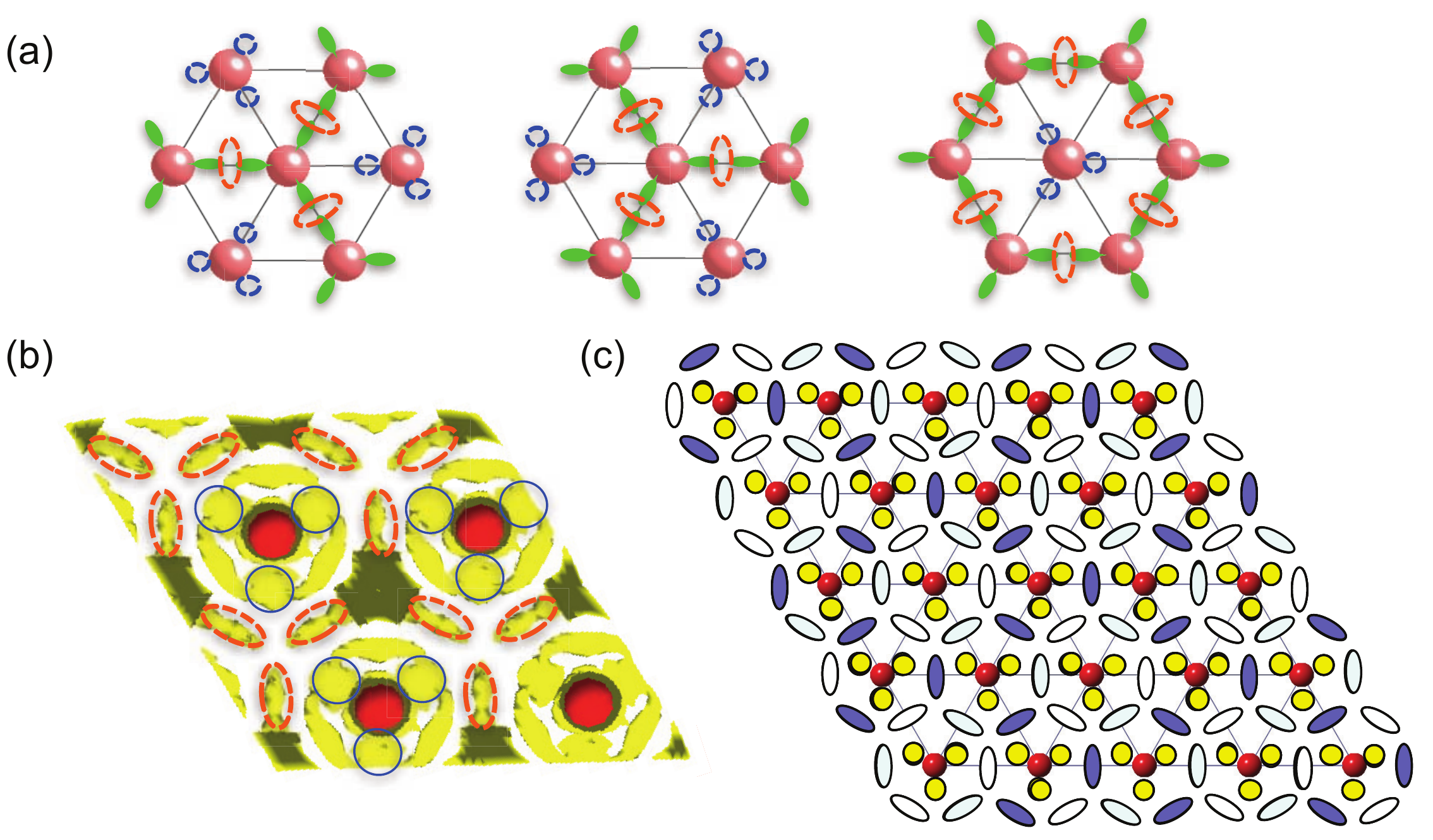}
\end{center}
\caption{\label{fig-Se1conjugate} (a) The three degenerate and entangled configurations of \(\pi\)-bond electrons (dashed orange ovals) and unpaired electrons (dashed blue circles) for the Se1 layer. The three \(\pi\)-bond electron clouds form a trimer, where the first two configurations can be viewed as TRS--protected topological orders and the third one is obtained via translation. (b) An electron density map cut for the Se1 layer obtained from the inverse Fourier transform of the diffraction data, where unexpected electron clouds (in dashed oval circles) among Se1-Se1 atoms are proposed to come from the \(\pi\)-bond electron contribution reflected by the time-average data of x-ray diffraction. (c) An integrated simulation for the electron density mapping of the Se1 layer made by using the  three entangled configurations shown in panel (a), which is actually composed of infinite replicas under quantum fluctuation but protected by  time-reversal symmetry in real space. [Panel (b) is reproduced from Ref. [\onlinecite{shu2018dynamic}] with permission; copyright © 2018, APS.]}
\end{figure}

Several valence-bond models for Bi\(_2\)Se\(_3\) have been proposed since 1958,\cite{Drabble:1958tf, Bhide:1971wm} but none of the proposed orbital hybridizations have been capable, until recently, of allowing a discussion of the surface properties.\cite{shu2018dynamic}  In this discussion, valence bonds are analyzed via octahedral coordination for all five elements in each quintuple layer with \(sp^3d^2\) hybridization for the valence-shell electrons, as shown in Fig.~\ref{fig-Bi2Se3}.  The electron configurations for Bi ([Xe]\(4f^{14}5d^{10}6s^26p^3\)) and Se ([Ar]\(3d^{10}4s^24p^4\)) having octahedral coordination prefers to form \(\sigma\) bonds via \(sp^3d^2\) hybridization for all atoms in each quintuple layer.  The five electrons in the valence shell for Bi suggest that  limited charge transfer is available within each quintuple layer without crossing the van der Waals gap in 3D, which is consistent with a narrow bulk band gap of \(
{\sim}0.3\) eV.  In addition, the remaining three unpaired electrons per Se1 facing the van der Waals gap or crystal surface after cleaving must seek  additional chemical bonding for further stability.  We  proposed three possible valence-bond arrangements for the three dangling \(sp^3d^2\) lobes as a trimer to form \(\pi\) bonds via side-to-side orbital overlap on the Se1-layer surface, as shown in Figs.~\ref{fig-Se1conjugate}(a)--\ref{fig-Se1conjugate}(c).  Because there are six identical Se1 atoms but only three unpaired \(sp^3d^2\) electrons per Se1 in one plane, the \(\pi\)-bond system must form a conjugated system because of sharing, similar to the case of graphene with one \(\pi\) bond shared by three neighboring C-C pairs at any instance.  

Unlike the \(\sigma\) bonds that do not change the crystal symmetry, a lattice distortion is expected for shared \(\pi\) bonds that induce a Peierls condensation to break the local crystal symmetry. However, the broken symmetry from three shorter bonds at any instance could be absorbed by a supersymmetry that contains both distortion and restoration in resonance with the choice of a larger basis.  The TRS-protected topological orders can be understood by the randomly selected \(\pi\)-bond trimer configuration in either right-hand chirality  or left-hand chirality  under quantum fluctuation; i.e.,  no difference occurs if a pattern is observed, like a film played  backward.  Conversely, when a ferromagnetic impurity sits near Se1, the spins of the unpaired electrons are expected to be polarized by the local magnetic field and to be pinned in either chirality [see Fig.~\ref{fig-chiral}(d)], and the TRS is broken.  A gap is opened after the TRS is broken, because no supersymmetry to absorb the distortion from the entangled condensation in real space, and the system falls back to the topologically trivial state without the required symmetry protection. 

Finally, the \(\pi\)-bond electrons can also be mapped similar to the \(\sigma\)-bond electrons as an electron crystal but, unlike the \(\sigma\)-bond electrons that share identical symmetry to the crystal structure, the \(\pi\)-bond electrons not only escape from the radar of  conventional band calculations due to additional instantaneous crystal-symmetry breaking  from the entangled and distorted condensation. Instead, they also constitute a conjugated system in dynamic form, which must include the discussion of entropy. In other words, the Gibbs free energy must be considered beyond the limitations of traditional band calculations which consider enthalpy reduction under the assumption of fixed crystal symmetry, at most using an additional term of SOC to relax the rule of band filling.  A good example is the electron density mapping of the Se1-layer obtained from the inverse Fourier transform of integrated diffraction data from x-ray diffraction experiments,\cite{shu2018dynamic} as shown in Fig.~\ref{fig-Se1conjugate}(b).  The additional electron clouds among the Se1 atoms cannot come from any \(\sigma\) bond but are assigned to \(\pi\)-bond electrons, and the integrated pattern shown in Fig.~\ref{fig-Se1conjugate}(c) being composed of the three possible orders shown in Fig.~\ref{fig-Se1conjugate}(a) agrees with the electron density map satisfactorily.

\subsection{Pb\(_{1-x}\)Sn\(_x\)Se: Crystalline-symmetry--protected topological insulator state}

\begin{figure}
\begin{center}
\includegraphics[width=6.5in]{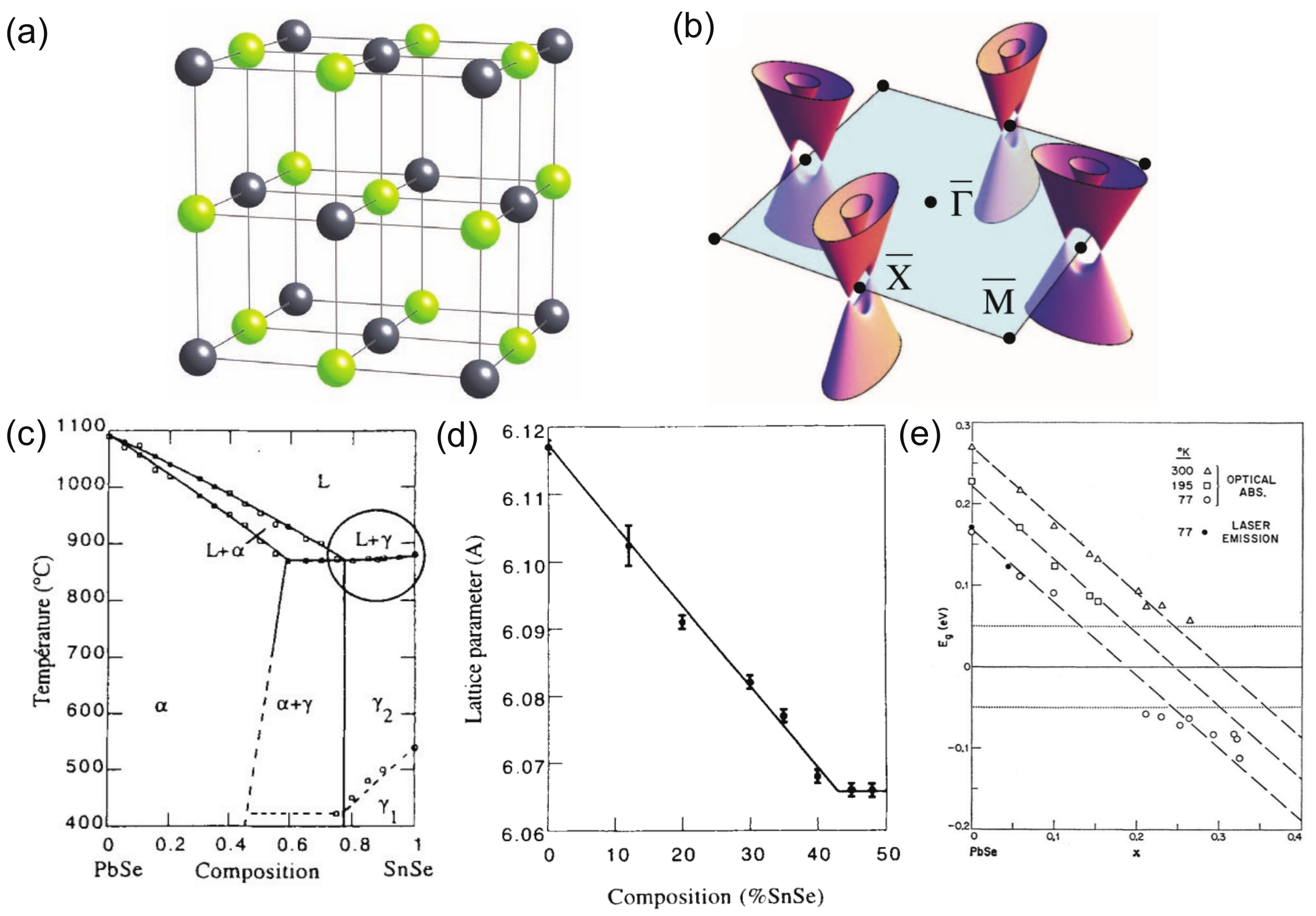}
\end{center}
\caption{\label{fig-PbSnSeBand} (a) The crystal structure of Pb\(_{1-x}\)Sn\(_x\)Se has the rock-salt-type cubic  space group \(Fm\bar{3}m\). (b) The surface Brillouin zone has pairs of Dirac nodes across the \(\bar{X}\) point in momentum space. Panels (c)--(e) show the PbSe-SnSe binary phase diagram, lattice parameters, and the bulk band gap of Pb\(_{1-x}\)Sn\(_x\)Se as a function of \(x\).  The \(L+\alpha\) mixture below \(x\sim0.4\) in the phase diagram suggests that strain energy is stored in the system. (Reproduced from Refs. [\onlinecite{okada2013observation, Corso:1995ky, STRAUSS:1967dx}] with permission; copyright © 2013, AAAS; copyright © 1995, ASM International;  copyright © 2018, APS.)  }
\end{figure}

Pb\(_{1-x}\)Sn\(_x\)Se (PSS) has been proposed as the first example of a TI with crystalline-symmetry--protected topological orders.\cite{Fu:2011ia, okada2013observation, Xu:2012bma, Dziawa:2012hx} The TI state of PSS is attributed to the cubic symmetry of rock-salt type [Fig.~\ref{fig-PbSnSeBand}(a)] that protects topological orders as a topological crystalline insulator (TCI).  The PbSe-SnSe binary is fully soluble for  \(x\lesssim0.4\) with linearly decreasing lattice parameters following Vegard's law because of the substitution of the smaller Sn into Pb sites [Fig.~\ref{fig-PbSnSeBand}(c)-(d)], which leads to the evolution from PbSe of rock-salt--type structure to SnSe of B29-type orthorhombic structure  with a similar valence-shell electron distribution in the heavy elements.\cite{STRAUSS:1967dx}  The size difference between Sn and Pb introduces compressive strain for the Pb\(_{1-x}\)Sn\(_x\)Se alloy system, until the SnSe of the  orthorhombic layered structure (space group \(Pbnm\)) is formed with severe Sn vacancies.  Band inversion  occurs near \(x\sim0.23\) and the material is confirmed to be a TCI to separate the narrow-gap semiconducting and the narrow-band-overlap semimetal states [Fig.~\ref{fig-PbSnSeBand}(e)].\cite{Dziawa:2012hx}  A similar TCI property has also been confirmed in the PbTe-SnTe alloy system of Pb\(_{1-x}\)Sn\(_x\)Te in the inverted-band regime for \(x\sim0.4\).\cite{Xu:2012bma}  In contrast to the TI materials discussed so far, PSS shows a pair of Dirac cones straddling the \(\bar{X}\) point for the (001) surface-state projection [Fig.~\ref{fig-PbSnSeBand}(b)], instead of a single Dirac cone, which was attributed to the Lifshitz transition as a result of (110) mirror-symmetry-protected topological orders.\cite{okada2013observation}   From the materials point of view, it is also interesting and puzzling to notice that PSS with a fcc cubic lattice does not cleave along the (111) planes that are in 2D close packing, but cleave easily to show (001) planes instead, which implies that a relatively stronger chemical bonding exists within the (001) planes. 

\begin{figure}
\begin{center}
\includegraphics[width=4.5in]{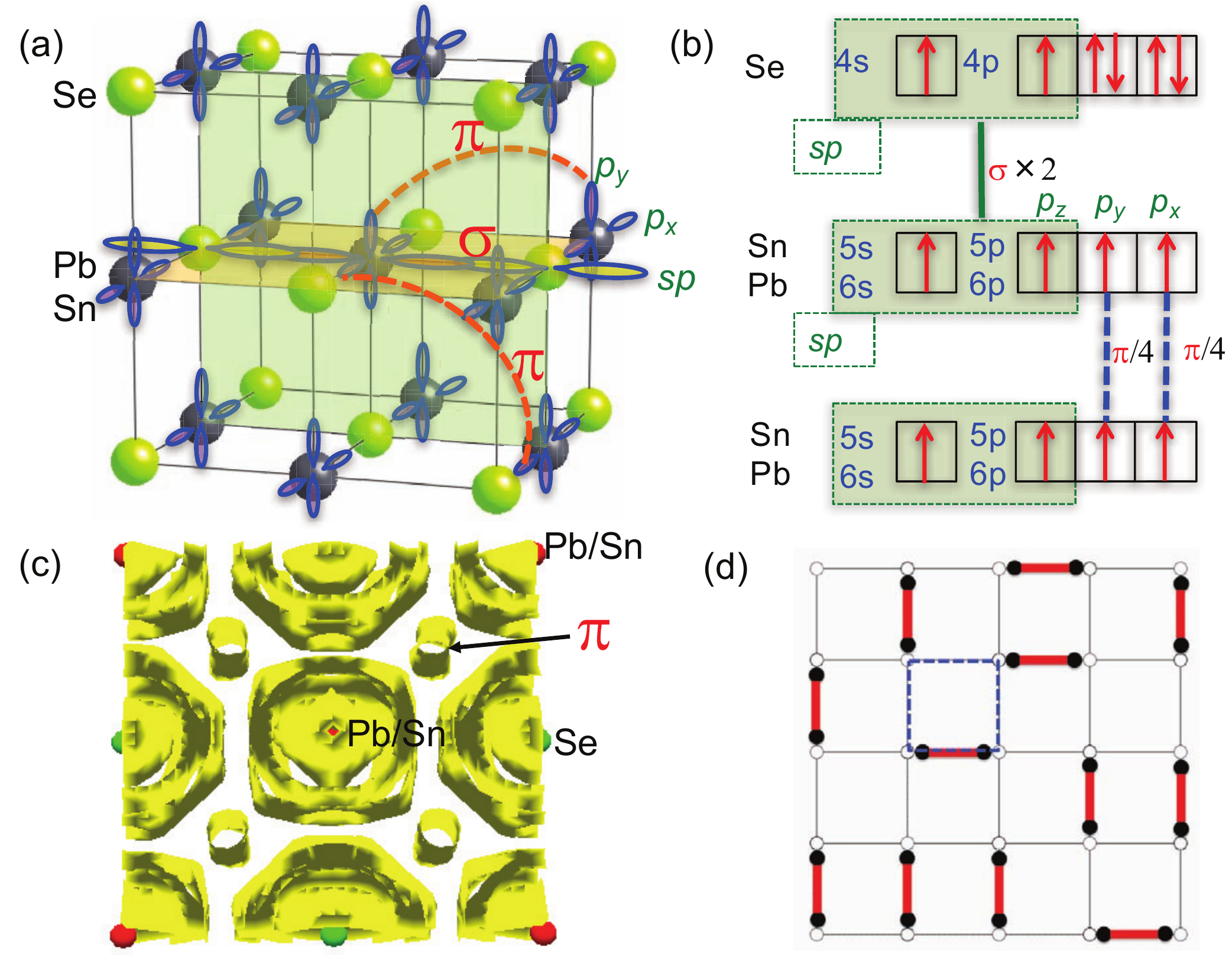}
\end{center}
\caption{\label{fig-PbSeVB}  (a), (b) The cubic crystal structure of Pb\(_{1-x}\)Sn\(_x\)Se is built by the \(\sigma\) bonds from the orbital overlap of Pb(Sn)-Se  \(sp\) hybrid orbitals in chain, and two conjugated \(\pi\)-bond systems in orthogonal planes among Pb(Sn) using unpaired electrons in \(p_{x,y}\) of Pb(Sn) separately, which are shared by four neighboring Pb(Sn) atoms in each square plane of 3D. (c) The electron density mapping for a single layer cut from the (001) plane, where the \(\pi\)-bond electron cloud is confirmed among Pb(Sn). (d) Consider the Peierls condensation due to \(\pi\)-bond formation, one of the infinite replicas for the Pb(Sn) sublattice is shown with \(\pi\)-bond electrons indicated by red bars, and the cubic-symmetry--protected (\(C_4\) square symmetry in 2D)  topological orders are revealed in real space.  (Reproduced from Ref. [\onlinecite{Shu:2015hk}] with permission; copyright AIP publishing.)  }
\end{figure}

Consider the evolution of the PbSe-SnSe binary phase diagram based on the integrated information including atom size, lattice parameters, melting points, and band inversion: the compressive strain driven by the Sn substitution should not be ignored, which is similar to the strain-driven TI states of \(\alpha\)-Sn and Bi\(_{1-x}\)Sb\(_x\), as shown above (Figs.~\ref{fig-alphaSn} and \ref{fig-BiSb}).  Starting from the atomic electron configurations of Pb ([Xe]\(4f^{14}5d^{10}6s^26p^2\)), Sn ([Kr]\(4d^{10}5s^25p^2\)), and Se ([Ar]\(3d^{10}4s^24p^4\)), the cubic structure for all atoms with six neighbors in octahedral coordination is proposed to have a valence-bond model, as shown in Fig.~\ref{fig-PbSeVB}.  Consider only one (001) plane: the four valence-shell electrons of Pb(Sn) and Se are hybridized into \(sp\) orbitals to form \(\sigma\) bonds bridging Pb(Sn)-Se in the \(z\) direction, and the two remaining unpaired \(p_{x,y}\) electrons can simultaneously form \(\pi\) bonds via side-to-side overlap in two orthogonal planes after the proper tuning of the strain energy and bond length.  For a sample of \(x\sim0.23\) with confirmed existence of a Dirac cone, electron density mapping for a single (001) layer has shown evidence of contributions from the \(\pi\)-bond electrons among Pb(Sn)-Pb(Sn), as shown in Fig.~\ref{fig-PbSeVB}(c).  Assuming that \(\sigma\) bonds are in the \(z\) direction, two \(\pi\) bonds shared by four neighboring Pb(Sn) atoms in each orthogonal plane are slightly above and below the \(xz\) and \(yz\) planes where \(\sigma\) bonds reside [Fig.~\ref{fig-PbSeVB}(a)].  Although the \(\pi\) bond is a conjugated system under quantum fluctuation with expected instantaneous distortion-restoration due to condensation, as illustrated by one of the infinite replicas shown in Fig.~\ref{fig-PbSeVB}(d), no clear symmetry breaking from the cubic symmetry (square symmetry in 2D) is found from the time-averaged diffraction data [Fig.~\ref{fig-PbSeVB}(c)], which is consistent with the theoretical prediction that these topological orders are protected by the crystalline symmetry, i.e., all real-space dynamic local distortions are absorbed by the cubic supersymmetry.  Note also that the pair of Dirac nodes being identified near the \(\bar{X}\) points are not just protected by the (110) mirror symmetry, which is actually coming from the side view of \(\pi\) bond electrons positioning slightly above and below all orthogonal (001) planes, i.e., coming from the side view of \(\pi\) bond electrons of (100) and (010). In other words, it would show similar double Dirac cones near the \(K\) point for graphene as well---if the graphene sample were flipped 90\(^\circ\) vertically in the ARPES experiment for the E(k) mapping.

\section{Topological Dirac semimetals}

A semimetal is defined for materials with finite band overlap (direct or indirect) between the conduction and valence bands for the bulk, and a Dirac semimetal denotes materials with the conduction and valence bands touching at a single Dirac node.  Conversely, topological Dirac semimetals are defined for materials with surface states in a pair of Fermi arcs that connect the two bulk Dirac nodes.\cite{Xu:2015kx}  Only a few topological Dirac semimetals have been confirmed by ARPES experiments, including Na\(_3\)Bi and Cd\(_3\)As\(_2\).\cite{liu2014discovery, neupane2014observation}  

\subsection{Na\(_3\)Bi}

\begin{figure}
\begin{center}
\includegraphics[width=5.5in]{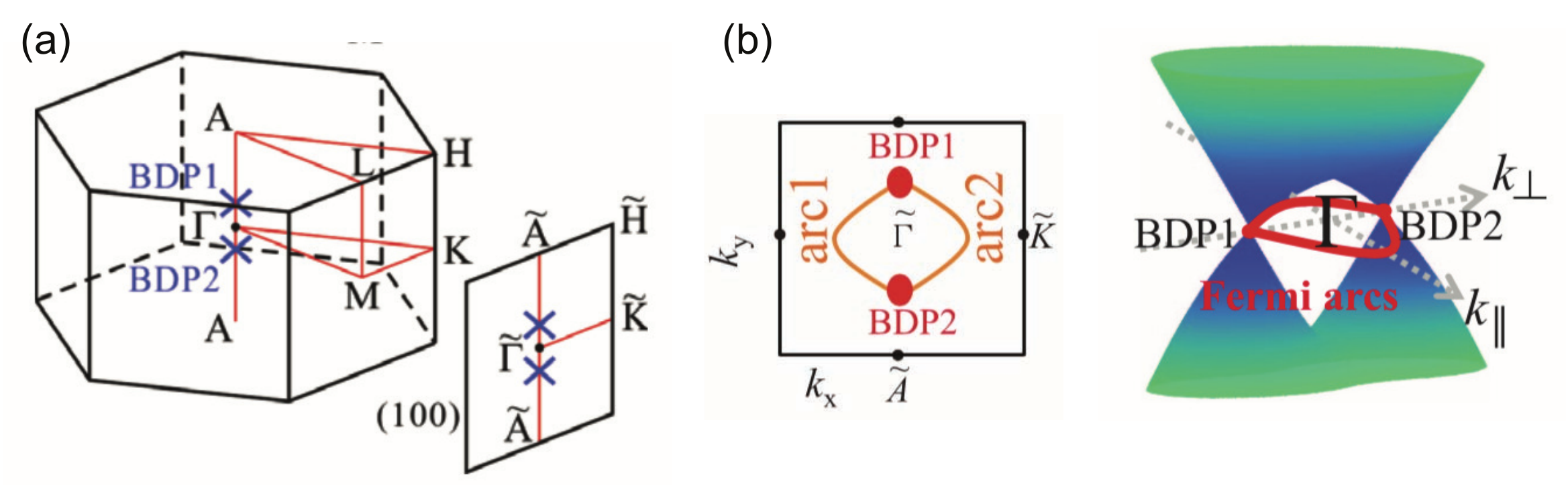}
\end{center}
\caption{\label{fig-Na3BiDirac} (a) The bulk Brillouin zone of Na\(_3\)Bi of space group \(P6_3/mmc\) having layers of Na-Bi honeycomb lattice with apical Na above and below each Bi, which is shown to have two Dirac nodes along the \(\Gamma\)-\(A\) direction. (b) The two Dirac nodes are connected by the two Fermi arcs of the surface state at the Fermi level. (Reproduced from Ref. [\onlinecite{Xu:2015kx}] with permission; copyright © 2015, American Association for the Advancement of Science.) }
\end{figure}

Na\(_3\)Bi has been confirmed by ARPES experiments to be a topological Dirac semimetal  showing both bulk Dirac nodes and the connecting Fermi arc pair for the surface state at the Fermi level, as shown in Fig.~\ref{fig-Na3BiDirac}.\cite{Xu:2015kx, Wang:cmANUFtU}   For a topologically trivial Dirac semimetal, a bulk Dirac node represents a single point touching between the conduction and valence bands, so that 3D metallic conduction is possible for the valence electrons moving to the conduction band in a specific direction defined by the \(k\) point of the Dirac node.  Conversely, for a topological Dirac semimetal such as Na\(_3\)Bi,  two Dirac nodes are connected by two one-way Fermi arcs for the spin-polarized surface electrons at the Fermi level. The question remains on  how to interpret such a band picture in real space.

\begin{figure}
\begin{center}
\includegraphics[width=5.5in]{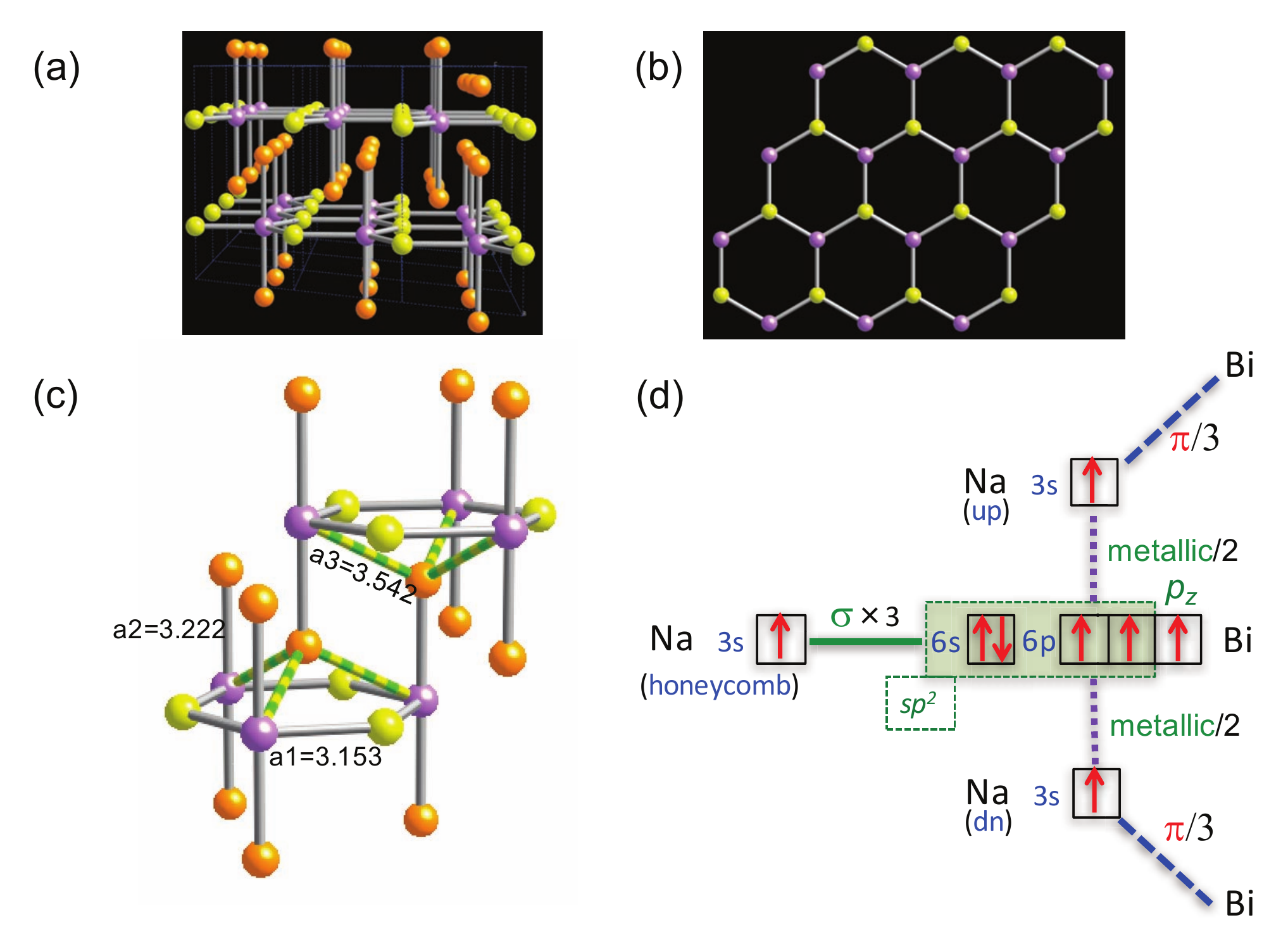}
\end{center}
\caption{\label{fig-Na3BiVB} (a), (b) The crystal structure of Na\(_3\)Bi is composed of a Bi-Na honeycomb lattice with apical Na (orange color) above and below each Bi.  (c) There are three different Na-Bi bond lengths: \(a_1=3.153\) \AA\ in the honeycomb layer, \(a_2=3.222\) \AA\ between Bi and the upper and lower Na, and the interlayer \(a_3=3.542\) \AA. (d) The proposed valence-bond model for Na\(_3\)Bi, including \(a_1\) corresponds to the three \(\sigma\) bonds for the honeycomb lattice, \(a_2\) corresponds to the metallic bonding of electron sharing between the two Na above and below Bi in resonance, and \(a_3\) corresponds to the \(\pi\) bond being shared by the three Bi in the neighboring layer as a conjugated system.  }
\end{figure}

The crystal structure of Na\(_3\)Bi is actually a derivative of graphite composed of 2D (BiNa)\(_n\) honeycomb lattices coupled to a Na double layer, as shown in Fig.~\ref{fig-Na3BiDirac}(a).  There are three slightly different bond lengths for Bi-Na based on the refined crystal structure,\cite{Steigmann:1968kv} including \(a_1=3.153\) \AA\ among the nearest-neighboring Bi-Na in the honeycomb layer,  \(a_2=3.222\) \AA\ between Bi-Na for Na sitting directly above and below Bi, and  \(a_3=3.542\) \AA\ between Bi-Na for Na sitting above and below each Bi-Na honeycomb center.  Following the requirements of (i) \(\sigma\)-bond formation via \(sp^2\) orbital hybridization for Bi in each honeycomb lattice and (ii) the strongest bond strength implying  the shortest bond length, a valence-bond model is proposed for Na\(_3\)Bi (Fig.~\ref{fig-Na3BiVB}).  The five valence-shell electrons of Bi ([Xe]\(4f^{14}5d^{10}6s^26p^3\)) are likely to distribute three valence electrons into a hybridized \(sp^2\) orbital to form three \(\sigma\) bonds with Na in a honeycomb lattice, as shown by the shortest bond length \(a_1\).  For the two remaining valence electrons in Bi-6\(p\), one is likely to share an electron with two apical Na-3\(s\) orbitals in metallic bonding of limited freedom, as reflected by the intermediate \(a_2\). The remaining unpaired Na-3\(s\) electron can form a conjugated \(\pi\)-bond system with the three Bi-6\(p_z\) orbitals in the neighboring layer, as suggested by the weakest bonding from the longest bond length \(a_3\).  Comparing the crystal structure with the first Brillouin zone shown in Fig.~\ref{fig-Na3BiDirac}, the two bulk Dirac nodes clearly correspond to the two layers of \(\sigma\)-bond electrons among Na(up)-Bi-Na(dn) along the \(\Gamma\)-\(A\) direction in the Brillouin zone. The electrons in the conjugated \(\pi\)-bond systems are used to bridge the bulk of the Bi-Na honeycomb layers via a quasi-2D electron conjugated \(\pi\)-bond network (\(a_3\)) and are connected by the Na(up)-Bi-Na(dn) channels (\(a_2\)) in resonance, which is consistent with the description of a band picture having Fermi arcs enter and leave the bulk Dirac node singularity, especially with a fixed chirality as a result of the one-way choice of \(\sigma\)-bond sharing among the Na-Bi-Na route.

\subsection{Cd\(_3\)As\(_2\)}

\begin{figure}
\begin{center}
\includegraphics[width=5.5in]{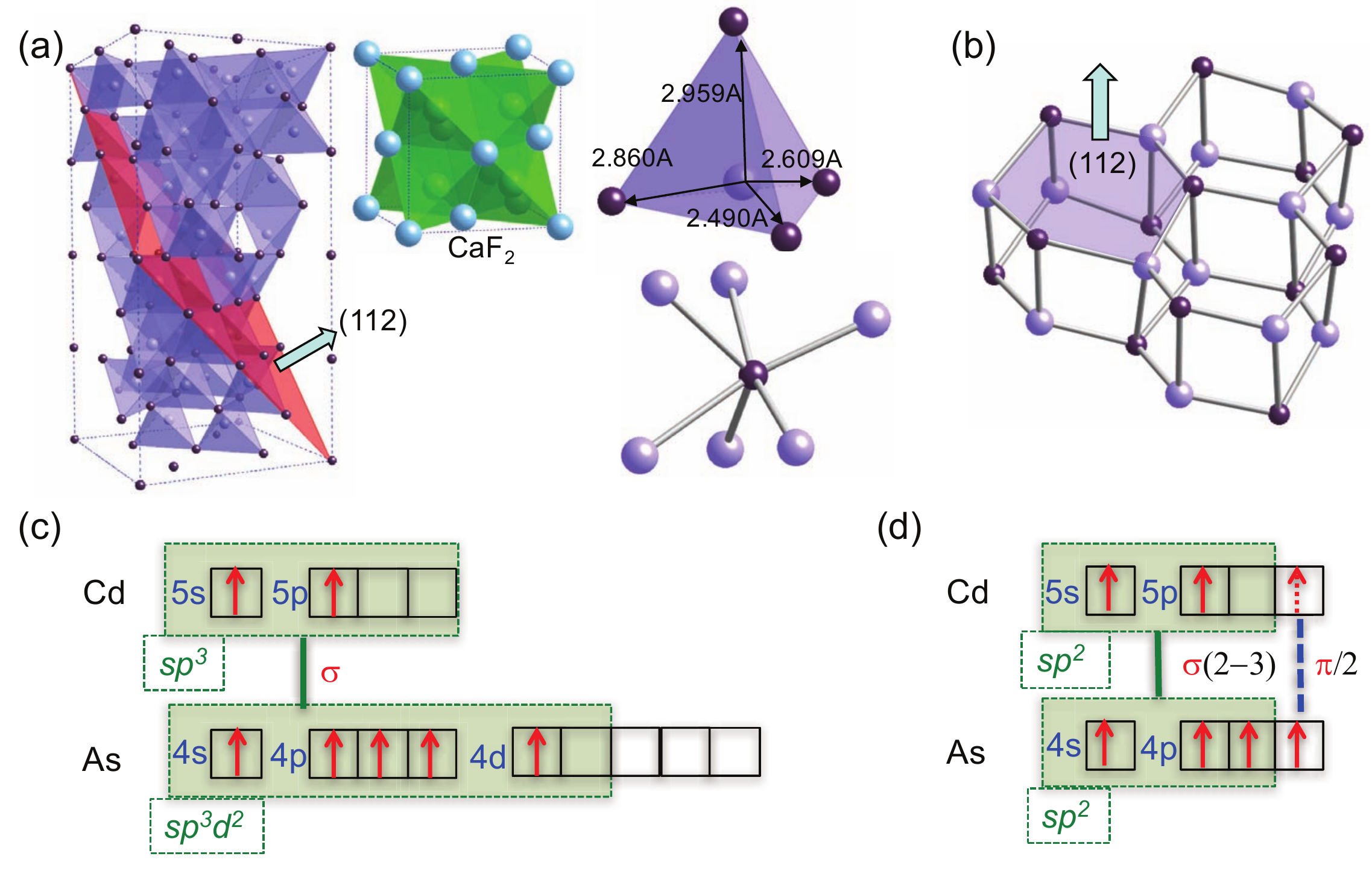}
\end{center}
\caption{\label{fig-Cd3As2} (a) The crystal structure of Cd\(_3\)As\(_2\) can be viewed as a defect-type fluorite structure (CaF\(_2\)) of systematic vacancies in a large unit cell.  The bond length and angle suggest severe distortion from the ideal \(\text{CdAs}_4\) tetrahedron or \(\text{AsCd}_6\) octahedron.\cite{Steigmann:1968kv} (b) The (112) plane shows a slightly distorted honeycomb lattice.  (c) The valence-bond models for Cd\(_3\)As\(_2\) as a semimetal, and (d) as a topological Dirac semimetal. To create a 2D conjugated \(\pi\)-bond system in Cd\(_3\)As\(_2\), the valence-bond model is likely to switch from the 3D \(\sigma\)-bond system formed with Cd-\(sp^3\) and As-\(sp^3d^2\) hybridized orbitals to quasi-2D \(\sigma\) bonding with \(sp^2\) hybridized orbitals for both, and a conjugated \(\pi\)-bond system becomes possible.   }
\end{figure}

As one of the very few confirmed topological Dirac semimetals, the band picture for Cd\(_3\)As\(_2\) is similar to that of Na\(_3\)Bi having two Dirac nodes at the Fermi level, and ultrahigh mobility has also been identified.\cite{liang2015ultrahigh, Guo:2016di, neupane2014observation}  Conversely, the space group assignment for Cd\(_3\)As\(_2\) has been under debate for a long time, mostly because the large unit cell of over 80 atoms makes it difficult to determine whether the space group is centrosymmetric (\(I4_1acd\) No.142) or noncentrosymmetric (\(I4_1cd\) No.110).\cite{Steigmann:1968kv, Arushanov:1980uj} However, recently Ali \textit{et al.} confirmed the centrosymmetric space group of \(I4_1acd\) (No. 142) with single-crystal diffraction.\cite{Ali:2014kc}  The crystal structure of Cd\(_3\)As\(_2\) should be viewed as a defect-type fluorite CaF\(_2\) structure (\(Fm\bar{3}m\) No. 225) with a slightly distorted Cd-As honeycomb lattice in the (112) direction, as shown in Fig.~\ref{fig-Cd3As2}(a).  The building block of the Cd\(_3\)As\(_2\) crystal is \(\text{CdAs}_4\)-tetrahedron, but it is greatly distorted from the ideal tetrahedron, as shown by the bond lengths in one of the \(\text{CdAs}_4\)-tetrahedra [Fig.~\ref{fig-Cd3As2}(a)], presumably due to the statistical arrangement of empty Cd sites along the body diagonal line within each defective fluorite unit.\cite{Arushanov:1980uj}  Conversely, it is interesting to find  the systematic change of symmetry identified by Ali \textit{et al.} from the high-temperature phase of \(Fm\bar{3}m\) (No. 225), to the intermediate-temperature phase of \(P4_2nmc\) (No.137), and to the low-temperature phase of \(I4_1/acd\) (No.142), which seems to violate the expected reduction of symmetry under reduced thermal fluctuation.  Such a violation could imply the emergence of a supersymmetry to protect the topological nature of the material, i.e., lower Gibbs free energy state is achieved via an increased entropy to absorb the fluctuating local distortion and restoration.  

Based on the atomic electron configuration and the coordination for both Cd ([Kr]\(4d^{10}5s^2\)) and As ([Ar]\(3d^{10}4s^24p^3\)), the coordination of the \(\text{CdAs}_4\)-tetrahedron expects \(sp^3\) hybridization for Cd and the \(\text{AsCd}_6\)-octahedron expects \(sp^3d^2\) hybridization for As. The valence-bond model for the bulk is shown in Fig.~\ref{fig-Cd3As2}(c).  However, since the Dirac cone is observed along the (112) direction from a slightly distorted honeycomb network, the valence electrons can be distributed among the three \(\sigma\) bonds in the honeycomb lattice via \(sp^2\) orbital hybridization, as illustrated in Fig.~\ref{fig-Cd3As2}(d).  Note that the fifth valence electron in the valence shell of As could  participate in either the 3D \(\sigma\)-bond formation or the 2D honeycomb \(\sigma\)-bond network with additional conjugated \(\pi\)-bond system.  The large unit cell and strain-induced local distortion makes it difficult to determine the exact crystal and band structures, but the current proposed valence-bond model can be used to improve our understanding of why Cd\(_3\)As\(_2\) is named a topological Dirac semimetal.  In an overly simplified description, the holes in the hybridized orbitals of Cd and As could be viewed as the origin of local charge transfer in 3D, but the conjugated \(\pi\)-bond system shown in the (112) plane becomes the middle ground for 2D-to-3D electron exchange, which corresponds to the band picture of a topological Dirac semimetal having 2D Fermi arcs interchangeably connecting the two Dirac nodes for the bulk at the Fermi level.\cite{neupane2014observation}

\section{Topological Weyl semimetals}

Theoretically, the topological Weyl semimetal is distinguished from the topological Dirac semimetal by the missing inversion or time-reversal symmetry of the former, so that the Dirac point splits into Weyl points.\cite{burkov2011aa}  Because Cd\(_3\)As\(_2\) has been confirmed to be a topological Dirac semimetal, it should have a centrosymmetric space group \(I4_1acd\) instead of the noncentrosymmetric space group \(I4_1cd\), as supported by the latest refinement by Ali \textit{et al.} using single-crystal samples.\cite{Ali:2014kc}  Because of the less stringent requirement on centrosymmetry, the potential of using topological Weyl semimetals in device application has been noticed.\cite{Hills:vz}  The first topological Weyl semimetal was proposed by Wan \textit{et al.} in the form of the  pyrochlore compound of Y\(_2\)Ir\(_2\)O\(_7\),\cite{Wan:2011hi} which was then verified experimentally in the binary systems  TaAs, TaP, NbS, and NbP.\cite{Xu:2015bt, Okada:2015jqa, yang2015weyl, Huang:2015ic}  

\subsection{TaAs}

\begin{figure}
\begin{center}
\includegraphics[width=5.5in]{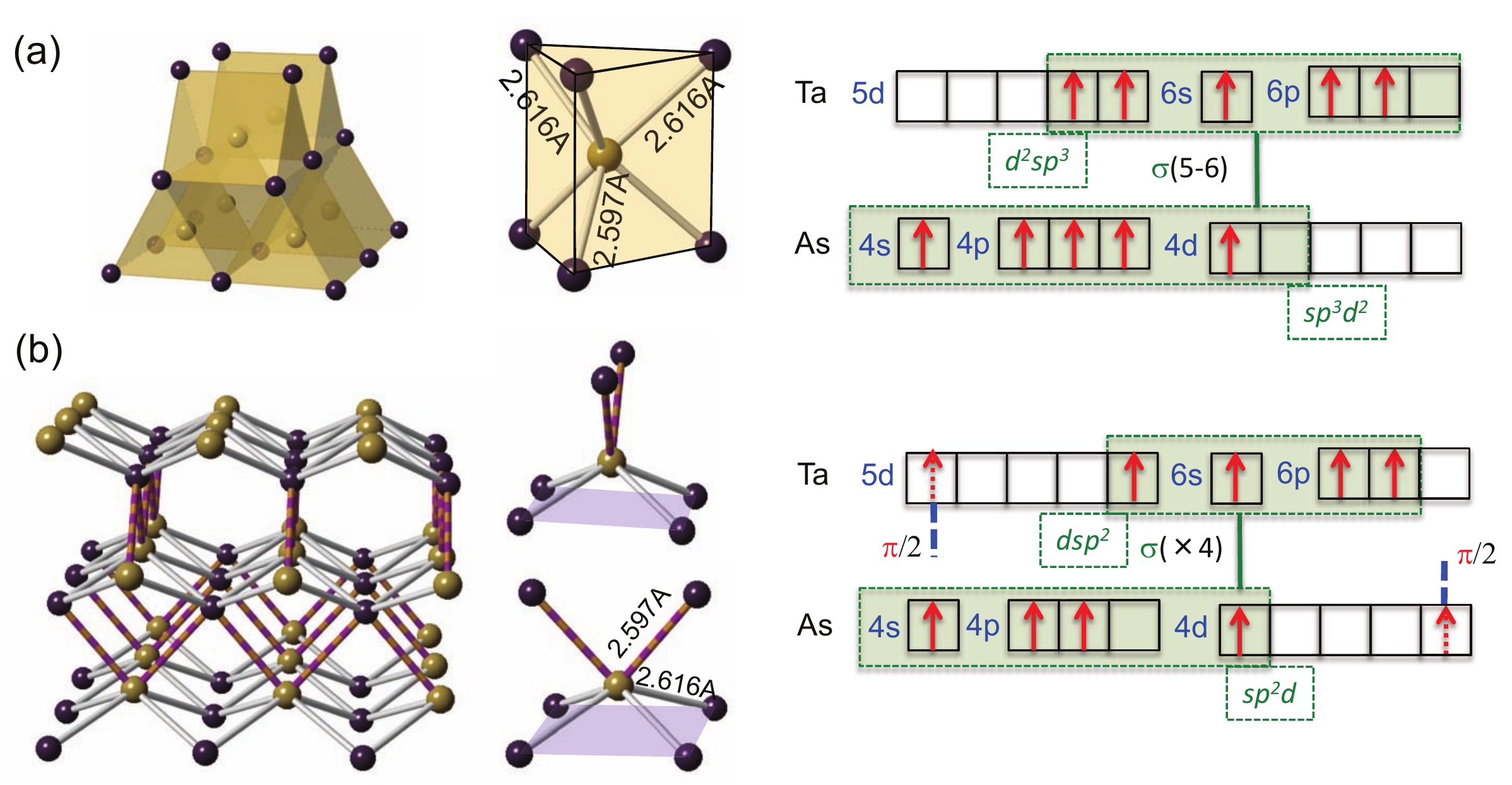}
\end{center}
\caption{\label{fig-TaAs} The crystal structure and valence-bond models for TaAs. (a) In view of the \(\text{TaAs}_6\) trigonal prismatic stacked in orthogonal arrangement, six \(\sigma\)-bonds are formed with Ta-\(d^2sp^3\) and As-\(sp^3d^2\) hybrid orbitals having bond lengths taken from Ref.~[\onlinecite{Boller:1963jja}].  (b) In view of  the \(\text{TaAs}_4\) square pyramid arrangement, four \(\sigma\) bonds are formed with Ta-\(dsp^2\) and As-\(sp^2d\) hybrid orbitals. An additional conjugated \(\pi\)-bond system becomes possible as a result of \(dd\)-type Ta-5\(d\) and As-4\(d\) side-to-side orbital overlap being shared by the two Ta-As of slightly shorter bond length (2.597~\AA). The Fermi arc connected splitted Dirac (Weyl) nodes in the band picture is reflected on the 90$^\circ$ turn plus glide operation of the shorter Ta-As bonds (dashed cylinder).  }
\end{figure}

The crystal structure of TaAs is tetragonal of space group \(I4_1md\), which lacks inversion symmetry, as expected for the requirement of a topological Weyl semimetal (Fig.~\ref{fig-TaAs}).  The conventional view for the crystal structure of TaAs is layers of \(\text{TaAs}_6\) or \(\text{AsTa}_6\) trigonal prismatic stacked at right angles between layers along the \(c\) direction of a tetragonal system.  However, upon examining the bond-length distribution among Ta-As,  one sees that two  types of valence-bond models are possible for the distribution of five electrons per valence shell for both Ta ([Xe]\(4f^{14}5d^36s^2\)) and As ([Ar]\(3d^{10}4s^24p^3\)).  If we distribute the five electrons in an trigonal prismatic coordination of six neighboring atoms, hybridized \(d^2sp^3\) orbitals for Ta and \(sp^3d^2\) orbitals for As are expected to satisfy the required six \(\sigma\) bonds [Fig.~\ref{fig-TaAs}(a)], but  one electron is missing in both hybridized orbitals, thereby suggesting further electron or hole sharing as a typical semimetal.  Conversely, the Ta-centered trigonal prismatic is actually composed of a square of longer bond (2.616~\AA) and a larger bond angle with two apices of a shorter bond (2.597~\AA), which strongly suggests an alternative valence-bond model of square planar coordinated hybridization of \(dsp^2\) for Ta and \(sp^2d\) for As to form four \(\sigma\) bonds.  In the model for square planar coordination,  note that the fifth valence electron per Ta and As would leave one of the $d$-orbitals (\(d_{x^2-y^2}\)) half filled in both [Fig.~\ref{fig-TaAs}(b)], which is expected to form a \(dd\)-type \(\pi\) bond among Ta-As in quasi-2D, similar to the \(pp\)-type \(\pi\) bond identified in graphene in 2D.   In particular, the \(\pi\) bond must be shared by the two apical As atoms of equal bond length (2.597~\AA) as a conjugated system under quantum fluctuation.  The two unpaired electrons and holes in these two valence-bond models are allowed to transfer charge between the 2D plane of the conjugated \(\pi\)-bond system and the \(\sigma\) bonds of the 3D bulk, which is consistent with the band picture showing chiral electrons of 2D states emerging from one Weyl point and sinking into the neighboring Weyl point via the designated path of Fermi arcs.\cite{Huang:2015ic}  The band picture for the spin-polarized description of a topological Weyl semimetal is complex, but the real-space view provides a more intuitive understanding that these complex band pictures are actually a description of the electron transfer between the 2D conjugated \(\pi\)-bond system in the square-coordinated plane and the 3D \(\sigma\)-bond system in a trigonal prismatic-coordinated bulk in resonance. In particular, there is always a \(90^\circ\) turn between neighboring trigonal prismatic layers in conjunction with the 2D square containing the conjugated \(\pi\)-bond system that requires a translation operation, as shown by the glide operation in the \(xy\) plane, which is described consistently by the momentum transfer showing Weyl points at different \(k\) points for a typical topological Weyl semimetal with split Dirac nodes.

\subsection{Y\(_2\)Ir\(_2\)O\(_7\)}

\begin{figure}
\begin{center}
\includegraphics[width=5.5in]{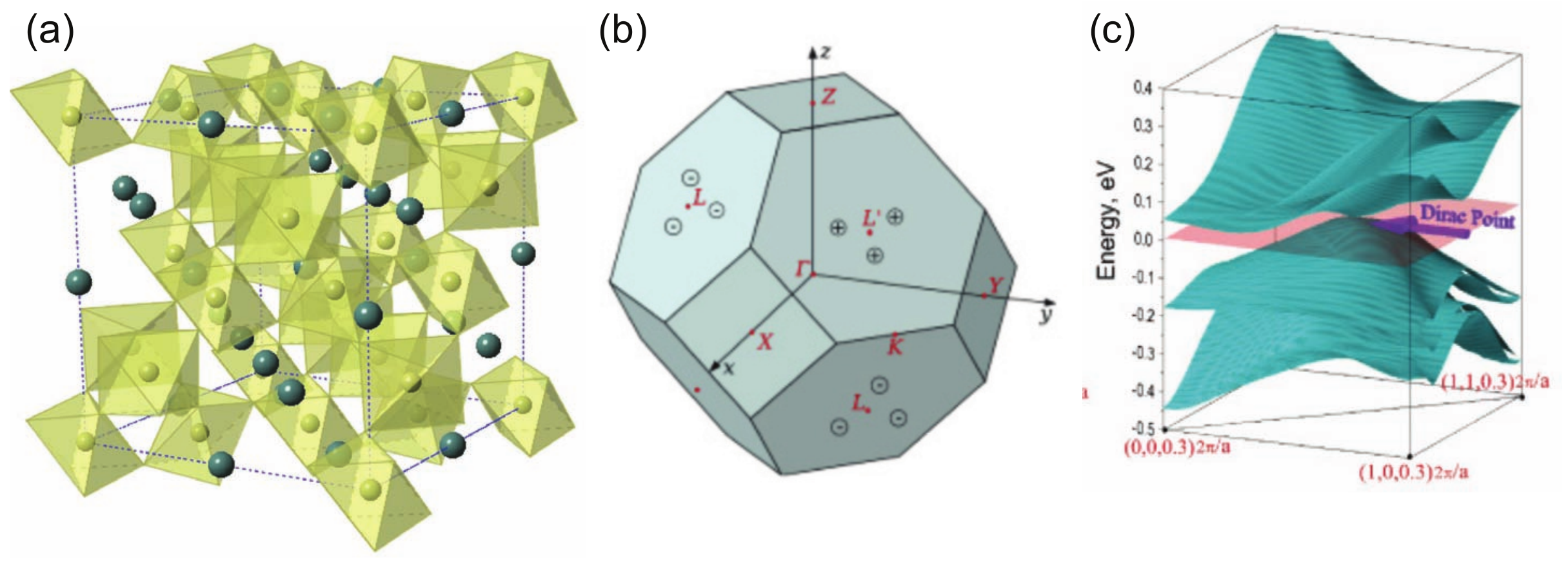}
\end{center}
\caption{\label{fig-pyrochlore} (a) The crystal structure of pyrochlore iridate Y\(_2\)Ir\(_2\)O\(_7\) has the space group  \(Fd\bar{3}m\) of cubic symmetry. (b) The Weyl points are indicated by the sign symbols shown on the surface of the first Brillouin zone. (c) Energy bands in the plane of \(k_z=0.6\pi/a\) to show a predicted Weyl point at the Fermi level. (Reproduced from Ref. [\onlinecite{Wan:2011hi}] with permission; copyright © 2018, APS.)   } 
\end{figure}

\begin{figure}
\begin{center}
\includegraphics[width=5.5in]{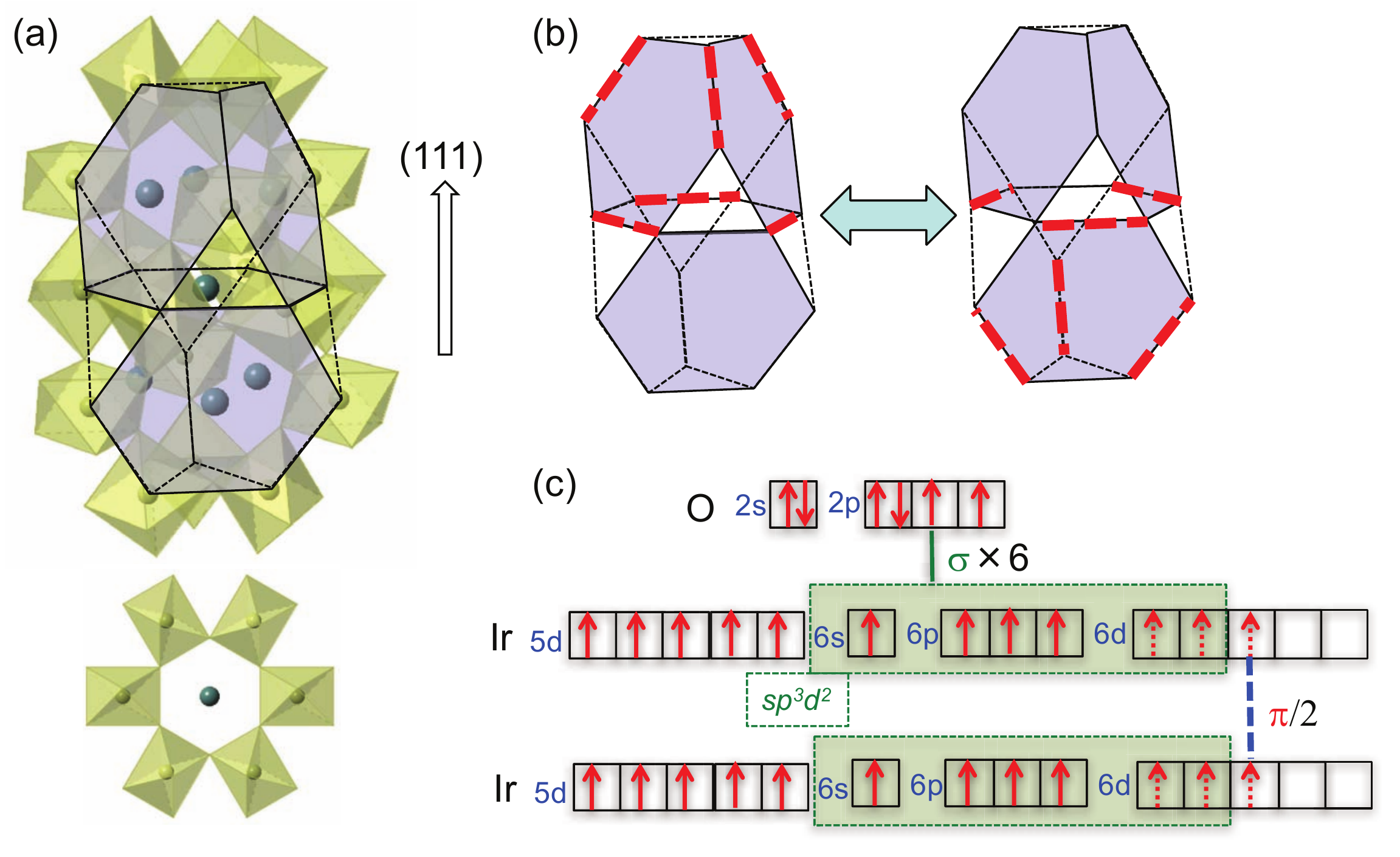}
\end{center}
\caption{\label{fig-YIOpi} (a) The building block of Y\(_2\)Ir\(_2\)O\(_7\) is the honeycomb ring composed of \(\text{IrO}_6\) octahedra with Y in each honeycomb ring center. (b)  A conjugated \(\pi\)-bond (red dashed line) system is formed among Ir using one of the half-filled 6\(d\) orbitals. (c) The valence-bond model of Y\(_2\)Ir\(_2\)O\(_7\) with three electrons donated from Y to Ir.  }
\end{figure}

When Wan \textit{et al.} first proposed the topological Weyl semimetal state in the pyrochlore iridate Y\(_2\)Ir\(_2\)O\(_7\) via \(\text{local density approximation}+U\) calculations,\cite{Wan:2011hi} it was truly difficult to realize that it could be viewed as a 3D analog of graphene from the band picture alone.  Based on the understanding of the valence-bond model of graphene and the successful modeling of topological materials discussed so far,  applying the same valence-bond approach is desirable to interpret why and how 2D electron states coexist with the 3D bulk state via symmetry-protected topological orders for Y\(_2\)Ir\(_2\)O\(_7\).   The crystal structure of Y\(_2\)Ir\(_2\)O\(_7\) is presented in Fig.~\ref{fig-pyrochlore}(a), with corner-sharing \(\text{IrO}_6\)-octahedra and Y in the interstitial sites  arranged in the cubic symmetry of space group \(Fd\bar{3}m\).  The view along the (111) direction of close-packing plane helps to reveal that the actual building block is a honeycomb ring composed of corner-sharing \(\text{IrO}_6\)-octahedra with Y in the honeycomb center. Seven honeycomb rings form two inverted pyramids along the (111) direction without inversion symmetry, as shown in Fig.~\ref{fig-YIOpi}(a).  The Brillouin zone with three Weyl nodes along the \(\Gamma\)-\(L\) direction are proposed by the calculations, as shown in Fig.~\ref{fig-pyrochlore}(b).

The valence-bond model of Y\(_2\)Ir\(_2\)O\(_7\) can be constructed based on the atomic electron configurations of Y ([Kr]\(4d^15s^2\)), Ir ([Xe]\(4f^{14}5d^76s^2\)), and O ([He]\(2s^22p^4\)), and the honeycomb rings are composed of corner-sharing \(\text{IrO}_6\)-octahedra and interstitial Y, as shown in Fig.~\ref{fig-YIOpi}(a).  Y donates three valence-shell electrons to the hybridized \(sp^3d\) orbital of Ir to allow six \(\sigma\)-bond formations with the neighboring O of unpaired \(2p\), and the Ir \(5d^5\) orbital has a half-filled special stability, so that one unpaired electron per Ir remains in the \(6d\) orbital.  Along the close-packing (111) direction, three edge-shared honeycomb rings are seen above and below the base honeycomb ring, as illustrated in Figs.~\ref{fig-YIOpi}(b).  The unpaired 6\(d\) electron per Ir could form side-to-side orbitals overlapping as  \(dd\)-type \(\pi\) bonds among the six Ir per honeycomb ring, which leads to a conjugated \(\pi\)-bond system under quantum fluctuation between the two cages, as illustrated by the six red dashed lines to represent \(\pi\) bonds among the twelve Ir atoms [Fig.~\ref{fig-YIOpi}(b)], which is similar to the situation of the 2D conjugated \(\pi\)-bond system for graphene having each \(\pi\) bond shared by three C-C pairs [Fig.~\ref{fig-eCrystal}(d)], this is why we might view Y\(_2\)Ir\(_2\)O\(_7\) to be a 3D analog of graphene.  Most importantly, the electrons in the \(sp^3d^2\) hybrid orbital for \(\sigma\)-bond formation and those in the \(\pi\) bond are interchangeable, and the exchange must follow the specific 2D \(\pi\)-bond path in clockwise or counterclockwise path of chiral nature, which is consistent with the band picture for Y\(_2\)Ir\(_2\)O\(_7\) as a topological Weyl semimetal having 2D electrons emerging from and entering the 3D bulk Fermi nodes via \(k\)-dependent spin-momentum-locked Fermi arcs.

\section{Nodal-line semimetals}

A nodal-line semimetal is distinguished from a topological Dirac semimetal and a topological Weyl semimetal by having the Fermi or Weyl nodes extended into a line (or a ring) in \(k\) space.  The typical nodal-line semimetal includes ZrSiS having Dirac nodal rings protected by nonsymmorphic symmetry,\cite{Neupane:2016cza} and PbTaSe\(_2\) having Weyl nodal rings protected by  reflection symmetry.\cite{Bian:2016bn}  In particular, PbTaSe\(_2\) has also been confirmed to be a superconductor, with \(T_c \sim3.8\)~K.\cite{Sankar:2017kqa}  On the surface, the crystal structures of both ZrSiS and PbTaSe\(_2\) seem to involve the van der Waals gap as  quasi-2D materials with quintuple and triple atomic layers, respectively; however, the calculated band picture strongly suggests otherwise. The implication of nodal lines with accompanied surface bands deserves a detailed analysis from the viewpoint of valence-bond symmetry.

\subsection{ZrSiS}

\begin{figure}
\begin{center}
\includegraphics[width=5.5in]{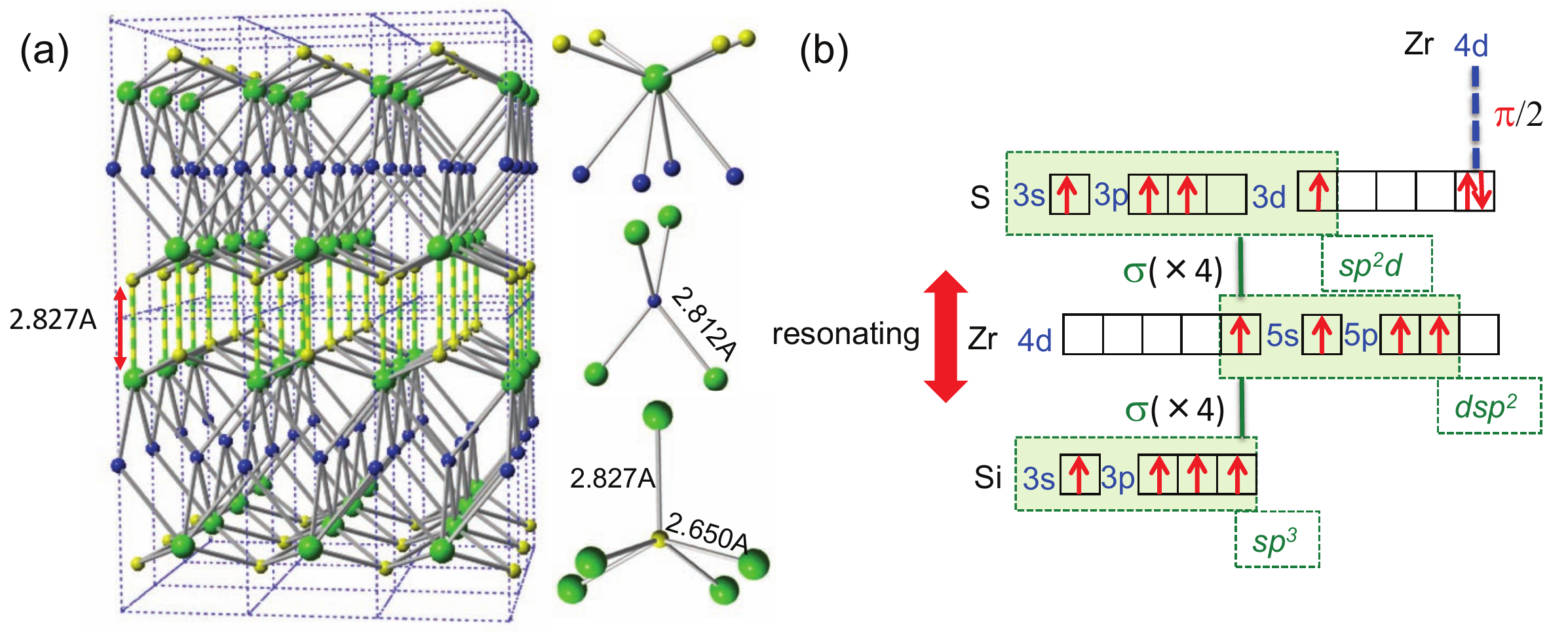}
\end{center}
\caption{\label{fig-ZrSiSVB} (a) The crystal structure of ZrSiS has a symmetry of space group \(P4/nmm\), which is a nonsymmorphic symmetry, as reflected on the glide and fourfold screw operation for \(\text{ZrS}_4\) and \(\text{SZr}_4\). (b) The covalent ZrSiS of \(\sigma\) bonds are formed with Zr \(dsp^2\), S \(sp^2d\), and Si \(sp^3\) hybrid orbitals based on the coordination.  The filled S 3\(d_{x^2-y^2}\) and empty Zr 4\(d_{x^2-y^2}\) could form a conjugated \(\pi\)-bond system to bridge ZrSiS quintuple layers. Note that the four valence electrons of Zr cannot satisfy both \(\text{ZrS}_4\) and \(\text{SiZ}r_4\) layers simultaneously but are forced to share in resonance as resonating valence bonds. }
\end{figure}

The crystal structure of ZrSiS is composed of S-Zr-Si-Zr-S stacked in quintuple layers, as shown in Fig.~\ref{fig-ZrSiSVB}.\cite{KleinHaneveld:1964bv}  Compared with  TI Bi\(_2\)Se\(_3\) having a similar quintuple layer with van der Waals gaps (Fig.~\ref{fig-Bi2Se3}), it is tempting to categorize the quintuple layers of ZrSiS as also having van der Walls gaps between S layers. However,  note that the Zr-S layers within each quintuple layer are particularly thin and the bond length of Zr-S across the quintuple layer gap is only slightly longer than those within the quintuple layer, which implies that an addition chemical bond emerges between  Zr-S across the quintuple layer gap.  The coordinations of Zr-Si-S are displayed in Fig.~\ref{fig-ZrSiSVB}(a), where the Zr layer is shared between the \(\text{SiZr}_4\) layer in tetrahedral coordination and the \(\text{ZrS}_4\) layer in a square pyramid coordination. Starting from the atomic electron configurations of Zr ([Kr]\(4d^25s^2\)), Si ([Ne]\(3s^23p^2\)), and S ([Ne]\(3s^23p^4\)), the four valence-shell electrons of Zr must be shared between the \(\text{SiZr}_4\) layer and the \(\text{ZrS}_4\) layer in resonance as resonating valence bonds, as illustrated by the proposed valence-bond model shown in Fig.~\ref{fig-ZrSiSVB}(b), where the four valence-shell electrons of Zr are distributed in half-filled \(dsp^2\) hybrid orbitals of square-pyramid coordination, the four electrons of Si are distributed in half-filled \(sp^3\) hybrid orbitals of tetrahedral coordination, and the six valence-shell electrons for S are distributed in half-filled square-coordinated \(sp^2d\) orbitals with two unpaired electrons remaining in 3\(d\) orbitals.  All \(\sigma\)-bond electrons among Zr-Si-S are reasonably satisfied to form a covalent quintuple layer, except for two unpaired electrons that remain in S 3\(d\) at the edge.  We propose that additional \(\pi\) bonds can be formed for  Zr-S by crossing the gap between quintuple layers via side-to-side orbital overlap of the empty Zr 4\(d_{x^2-y^2}\) and the filled S 3\(d_{x^2-y^2}\) orbitals.  Note that the \(\sigma\) bonds across the \(\text{SiZr}_4\) and \(\text{ZrS}_4\) units cannot be satisfied simultaneously and are forced to share in sync with the bridging \(\pi\)-bond system among the Zr-S across the quintuple layer, similar to the concept of a resonating valence bond, which was first proposed by L. Pauling for the discussion of Li metallic bonding and superconductivity,\cite{Pauling:1968te, Pauling:1968te} and later by  P. Anderson to explain the high-\(T_c\) superconductor of doped La\(_2\)CuO\(_4\).\cite{Anderson:1987iia}  Supporting evidence of such resonating \(\sigma\)-bond character is also hinted by the latest discovery that the Si surface can also be isolated via mechanical cleaving in addition to the expected S surface in the STM experiments.\cite{Su:2018vz}

The proposed \(dd\)-type \(\pi\) bond across the Zr-S gap (bond length 2.827~\AA) must be conjugated as a result of resonating \(\sigma\) bonds between \(\text{SiZr}_4\) and \(\text{ZrS}_4\) layers, because the two electrons in S 3\(d_{x^2-y^2}\), being responsible for the \(\pi\) bond, would preferably not stay  in the nonbonding state of S when the valence electrons of Zr are used in \(\text{SiZr}_4\) \(\sigma\) bonding.  The resonating \(\sigma\) bonds among \(\text{SiZr}_4\) and \(\text{ZrS}_4\) layers are consistent with the semimetal character predicted by the band calculation.  The 2D conjugated \(\pi\)-bond systems  exchange electrons with the bulk resonating \(\sigma\) bond to bridge not only the intralayer ZrSiS but also the quintuple layer across the gap.   Experimental observation for ZrSiS as a topological nodal-line semimetal has been confirmed with ARPES on  multiple pockets in various shapes with linearly dispersive surface states.\cite{Neupane:2016cza}  In addition, the  generation of nodal-line rings for ZrSiS by nonsymmorphic symmetry is attributed to the screw operation of the \(\text{ZrS}_4\) square pyramid, which was discussed in detail by Yan \textit{et al.} in the review about the symmetry of nodal-line materials.\cite{Yang:2018im} 


\subsection{PbTaSe\(_2\)}
\begin{figure}
\begin{center}
\includegraphics[width=5.5in]{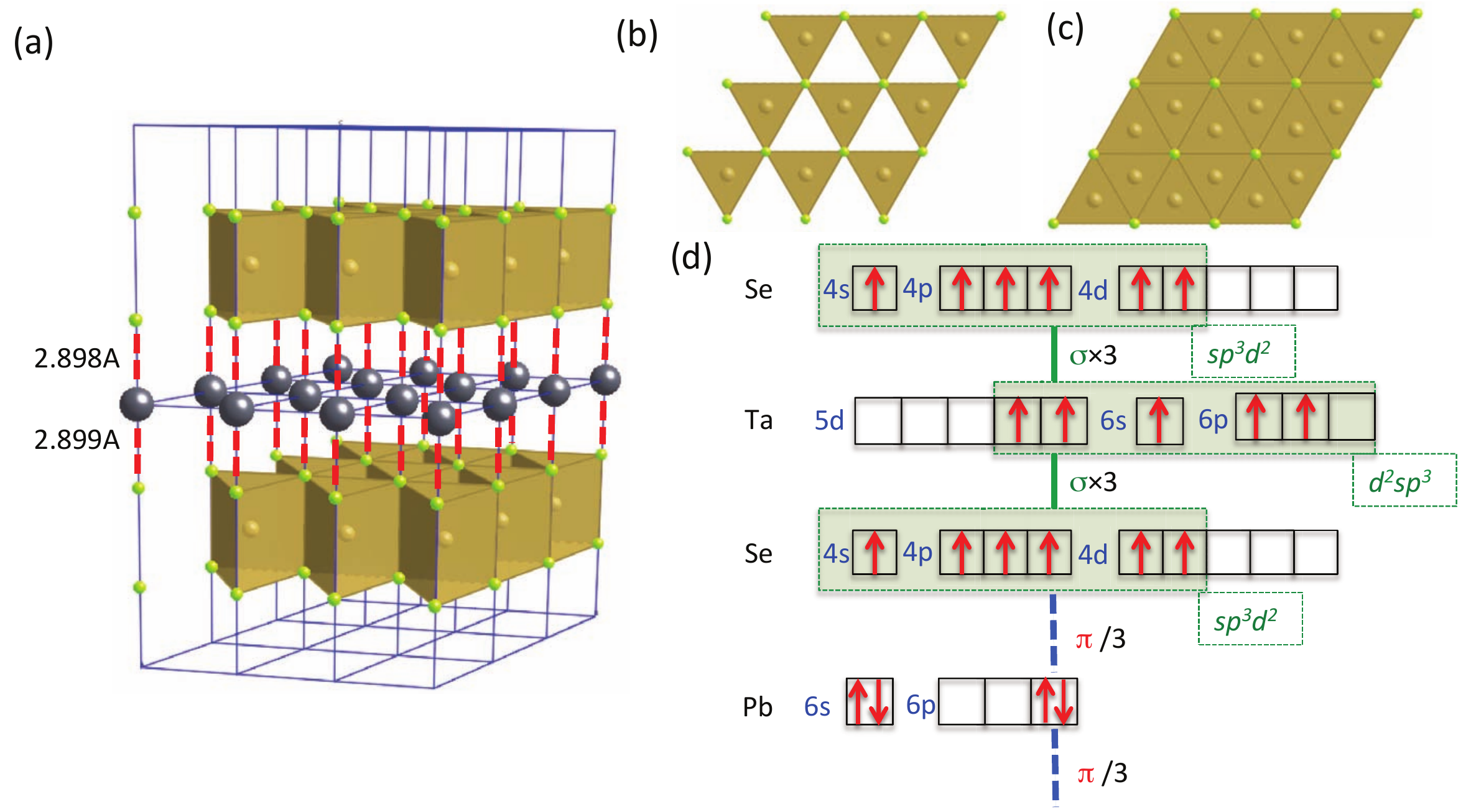}
\end{center}
\caption{\label{fig-PbTaSe2VB} (a) The crystal structure of PbTaSe\(_2\) can be viewed as a van der Walls material of TaSe\(_2\) with an intercalated Pb layer. (b), (c) Top view of TaSe\(_2\) layer to show that edge sharing of the \(\text{TaSe}_6\) trigonal prismatic of space group \(P\bar{6}m2\) is preferred, instead of \(P6/mmm\) with face-sharing trigonal prismatic of half occupancy for Ta.  (d) The proposed valence-bond model indicates that, in addition to the \(\sigma\)-bond formation with half-filled Ta-\(d^2sp^3\) and Se-\(sp^3d^2\) hybrid orbitals, an additional conjugated \(\pi\)-bond system is formed among the Se-Pb-Se layer across the original van der Walls gap of TaSe\(_2\). }
\end{figure}

PbTaSe\(_2\) can be viewed as a layered van der Walls material of TaSe\(_2\) with intercalated layers of Pb, as shown in Fig.~\ref{fig-PbTaSe2VB}(a).  Based on diffraction techniques alone, it is difficult to distinguish the two possible space groups of centrosymmetric \(P6/mmm\) (Fig.~\ref{fig-PbTaSe2VB}(c)) and noncentrosymmetric \(P\bar{6}m2\) (Fig.~\ref{fig-PbTaSe2VB}(b));\cite{Sankar:2017kqa} i.e., the former has half occupancy for Ta at (\(\frac{1}{3},\,\frac{2}{3},\,\frac{1}{2}\)) in \(P6/mmm\) of higher symmetry. The band calculations and ARPES results support the conclusion that PbTaSe\(_2\) is a Weyl nodal-line semimetal with broken centrosymmetry,\cite{Bian:2016bn} and the mirror symmetry for the Ta plane plays an essential role in protecting the topological nodal line.\cite{Yang:2018im}  Based on the trigonal prismatic coordination for Ta in the center, and the atomic electron configurations of Ta ([Xe]\(4f^{14}5d^36s^2\)), Se ([Ar]\(3d^{10}4s^24p^4\)), and Pb ([Xe]\(4f^{14}5d^{10}6s^26p^2\)), it is reasonable to hybridize the valence-shell electrons for both Ta and Se into \(d^2sp^3\) and \(sp^3d^2\), respectively, to form six required \(\sigma\) bonds per Ta, as shown in Fig.~\ref{fig-PbTaSe2VB}(d).   TaSe\(_2\) layer with one hole must induce electron sharing among the six expected \(\sigma\) bonds, which is consistent to the semimetal nature within each TaSe$_2$ layer.  

It is interesting to note that three unpaired electrons appear on both sides of the Se-Ta-Se triatomic layer per Se, similar to those shown on the surface of each quintuple layer in the typical topological insulator Bi$_2$Se$_3$ (see Fig.~\ref{fig-Bi2Se3}).  Since there are two unpaired electrons in 6\(p\) of Pb, one can be donated to Ta to satisfy the requirement of covalency for TaSe\(_2\) layer, and leave one unpaired electron in 6\(p_z\), which strongly suggests the possibility of \(\pi\)-bond formation between Pb 6\(p_z\) and the three half-filled \(sp^3d^2\) lobes of Se on either top or bottom side, as suggested by the slightly unequal bond length for Pb-Se shown in Fig.~\ref{fig-PbTaSe2VB}(a).  Alternatively, the two electrons in Pb 6\(p_z\) may both form \(\pi\) bond with the top and bottom Se simultaneously without donating to Ta to stabilize the TaSe$_2$ layer, however, the \(\pi\) bond between Pb and Se must still be shared among the three unpaired Se-$sp^3d^2$ lobes as a conjugated system.  The observed BCS type superconductivity could be closely related to the Pb rattling involved electron-phonon coupling for the Cooper pair exchange within and across layers.\cite{Sankar:2017kqa}  To understand the momentum translation among Weyl points for electrons moving from the bulk valence band to the conduction band following the designated 2D Fermi arcs,\cite{Bian:2016bn, Neupane:2016cza}  it can be translated into a real-space view of \(\pi\) bond electrons exchange between the \(\sigma\)-bonds in the bulk of \(\text{TaSe}_6\) octahedron and the conjugated quasi-2D \(\pi\)-bond system in Se-Pb-Se layer via the directional routes of \(\sigma\) and \(\pi\) bonds.

\section{Reflection and outlook}

Based on a valence-bond symmetry analysis,  the conjugated \(\pi\)-bond system identified in topological materials clearly constitutes a common signature in the real-space view, and its energy-momentum dispersion is described by the gapless Dirac cone showing the tunneling process of massless Dirac fermions.  The conjugated \(\pi\)-bond system in real space could be viewed as a supersymmetry system which is sufficient to absorb the instantaneous and spontaneous crystalline symmetry breaking and restoration, or as a thermodynamic system defined not only by its order parameters but also by its order indices. This is reflected by the additional dimensionality to include a topological classification in addition to the topologically trivial materials defined beforehand only by the rigid crystalline symmetry.  We believe the chemistry-physics divide can be removed once the understanding of topological materials is explored from both the real-space valence-bond model and from the reciprocal-space band pictures.  In particular, in the learning curve of topological materials from 1D to 3D, we find that the ground state of a thermodynamic system can no longer be limited to the lowest enthalpy of presumably fixed crystalline symmetry. The increased entropy from the conjugated \(\pi\)-bond system with constant breaking and restoration of the broken symmetry, i.e., the turned-on spin-orbital coupling (SOC) in band calculation, can also lead to the hidden ground state of topological materials, which is why more topological materials are waiting to be found beyond the traditional path of presumed rigid crystalline symmetry.

Following the valence-bond analysis for a series of prototype topological materials discussed above, from the simplest 1D \textit{trans}-PA to 2D graphene, and to the 3D TIs and semimetals containing electronic 2D edge states, the conjugated \(\pi\)-bond system as a common thread is identified.  The 1D or 2D \(\pi\)-bond conjugated system is attributed to the distortion and restoration of broken symmetry due to  electron pairing with  concomitant Peierls condensation.  These conducting edge states have real-space configurations of high entropy but can be simplified with a supersymmetry that absorbs the perturbation of the original broken crystalline symmetry.  Whereas the novel 2D electronic state is commonly found in the 3D bulk of topological materials, the observed Dirac-cone signature detected by  ARPES experiments should not be  attributed solely to the physical surface of a condensed-matter state, but should be interpreted as coming from the conjugated \(\pi\)-bond electrons confined within a plane in the 3D bulk in general.  From the materials-preparation perspective, besides picking out topological materials from the huge chemical database with a designed algorithm through symmetry and parity analyses, strain engineering at the interface of semimetal-semiconductor binaries  through the adjustment with substitution and epitaxial-growth techniques seems to be more accessible experimentally to harness the unique low-dimensional transport character of chiral electrons for device applications.     

\section*{Acknowledgments}

F.C.C. would like to thank G. J. Shu, Raman Sankar and M. Hayashi for many helpful discussions. The support provided by the Ministry of Science and Technology in Taiwan under project number MOST-106-2119-M-002 -035 -MY3 is acknowledged.   \\


\begin{thebibliography}{80}

\bibitem{Shirakawa:2006en} H. Shirakawa, Molecular Crystals and Liquid Crystals 171, 235 (1989).
\bibitem{FincherJr:1982hc} C. R. Fincher Jr., C. E. Chen, A. J. Heeger, A. G. MacDiarmid, and J. B. Hastings, Physical Review Letters 48, 100 (1982).
\bibitem{Geim:2011ed} A. Geim, Reviews of Modern Physics 83, 851 (2011).
\bibitem{Novoselov:2011gq} K. Novoselov, Reviews of Modern Physics 83, 837 (2011).
\bibitem{Haldane:2017et} F. D. M. Haldane, Reviews of Modern Physics 89, 040502 (2017).
\bibitem{Hasan:2010kua} M. Z. Hasan and C. Kane, Reviews of Modern Physics 82, 3045 (2010).
\bibitem{Bansil:2016bu} A. Bansil, H. Lin, and T. Das, Reviews of Modern Physics 88, 021004 (2016).
\bibitem{E:2018wp} E. Gibney, Nature 560, 151 (2018).
\bibitem{Hosoya:1975ea} H. Hosoya, K. Hosoi, and I. Gutman, Theoretica Chimica Acta 38, 37 (1975).
\bibitem{Dobrowolski:2003tq} J. C. Dobrowolski, Croatica Chemica Acta 76, 145 (2003).
\bibitem{Lifshitz:1960ux} I. M. Lifshitz, Soviet Physics JETP 11, 1130 (1960).
\bibitem{Neto:2009wq} A. C. Neto, F. Guinea, N. Peres, K. S. Novoselov, and A. K. Geim, Reviews of Modern Physics 81, 109 (2009).
\bibitem{Bethe:1931iz} H. Bethe, Zeitschrift Fur Physik 71, 205 (1931).
\bibitem{Affleck:1987te} I. Affleck, T. Kennedy, E. H. Lieb, and H. Tasaki, Phys. Rev. Lett. 59, 799 (1987).
\bibitem{Bonner:1964fm} J. C. Bonner and M. E. Fisher, Physical Review 135, A640 (1964).
\bibitem{Majumdar:1969iu} C. K. Majumdar and D. K. Ghosh, Journal of Mathematical Physics 10, 1388 (1969).
\bibitem{Haldane:1983wc} F. Haldane, Phys. Rev. Lett. 50, 1153 (1983).
\bibitem{Vasilev:2005tf} A. Vasil'ev, M. Markina, and E. Popova, Low Temperature Physics 31, 203 (2005).
\bibitem{BAERISWYL:1995ti} D. Baeriswyl and E. Jeckelmann, Materials Science Forum 191, 71 (1995).
\bibitem{Kivelson:2001js} S. A. Kivelson, Synthetic Metals 125, 99 (2001).
\bibitem{Belopolski:2017cp} I. Belopolski \textit{et al.}, Science Advances 3, e1501692 (2017).
\bibitem{Gutzler:2013ek} R. Gutzler and D. F. Perepichka, Journal of the American Chemical Society 135, 16585 (2013).
\bibitem{Fasolino:2007bz} A. Fasolino, J. H. Los, and M. I. Katsnelson, Nature Materials 6, 858 (2007).
\bibitem{Katsnelson:2007tu} M. I. Katsnelson, Materials Today 10, 20 (2007).
\bibitem{Wallace:1947dy} P. R. Wallace, Physical Review 71, 622 (1947).
\bibitem{Fuhrer:2010dk} M. S. Fuhrer, C. N. Lau, and A. H. Macdonald, MRS bulletin 35, 289 (2010).
\bibitem{Galeeva:2018fx} A. V. Galeeva, A. I. Artamkin, A. S. Kazakov, S. N. Danilov, S. A. Dvoretskiy, N. N. Mikhailov, L. I. Ryabova, and D. R.
Khokhlov, Beilstein Journal of Nanotechnology 9, 1035 (2018).
\bibitem{Moskvin:2008fo} P. Moskvin, V. Khodakovsky, L. Rashkovetskyi, and A. Stronski, Journal of Crystal Growth 310, 2617 (2008).
\bibitem{Bernevig:2006ij} B. A. Bernevig, T. L. Hughes, and S. C. Zhang, Science 314, 1757 (2006).
\bibitem{Konig:2007hs} M. Konig, S. Wiedmann, C. Brune, A. Roth, H. Buhmann, L. W. Molenkamp, X.-L. Qi, and S.-C. Zhang, Science 318, 766
(2007).
\bibitem{Rogalski:2005ju} A. Rogalski, Reports on Progress in Physics 68, 2267 (2005).
\bibitem{Kane:2005hl} C. Kane and E. J. Mele, Physical Review Letters 95, 226801 (2005).
\bibitem{Roman:1972fc} B. J. Roman and A. W. Ewald, Physical Review B 5, 3914 (1972).
\bibitem{Fu:2007ei} L. Fu and C. Kane, Physical Review B 76, 045302 (2007).
\bibitem{Bassani:1963kr} F. Bassani and L. Liu, Physical Review 132, 2047 (1963).
\bibitem{Butler:2013ha} S. Z. Butler \textit{et al.}, ACS Nano 7, 2898 (2013).
\bibitem{Lenoir:1998tl} B. Lenoir, A. Dauscher, M. Cassart, Y. I. Ravich, and H. Scherrer, Journal of Physics and Chemistry of Solids 59, 129
(1998).
\bibitem{Berger:1982ds} H. Berger, B. Christ, and J. Troschke, Crystal Research and Technology 17, 1233 (1982).
\bibitem{Ohtani:1994ix} H. Ohtani and K. Ishida, Journal of Electronic Materials 23, 747 (1994).
\bibitem{Cucka:1962jr} P. Cucka and C. S. Barrett, Acta Crystallographica 15, 865 (1962).
\bibitem{Hsieh:2008tf} D. Hsieh, D. Qian, L. Wray, Y. Q. Xia, Y. S. Hor, R. J. Cava, and M. Z. Hasan, Nature 452, 970 (2008).
\bibitem{Nakamura:2011gb} F. Nakamura \textit{et al.}, Physical Review B 84, 235308 (2011).
\bibitem{Yuki:2014ek} F. Yuki, O. Masao, and F. Hidetoshi, Journal of the Physical Society of Japan 84, 012001 (2014).
\bibitem{Hsieh:2009dp} D. Hsieh \textit{et al.}, Nature 460, 1101 (2009).
\bibitem{shu2018dynamic} G. J. Shu, S. C. Liou, S. K. Karna, R. Sankar, M. Hayashi, and F. C. Chou, Physical Review Materials 2, 044201 (2018).
\bibitem{Zhang:2009vu} H. Zhang, C. Liu, X. Qi, X. Dai, Z. Fang, and S. Zhang, 5, 438 (2009).
\bibitem{Drabble:1958tf} J. Drabble and C. Goodman, Journal of Physics and Chemistry of Solids 5, 142 (1958).
\bibitem{Bhide:1971wm} V. Bhide and B. Patki, Journal of Physics and Chemistry of Solids 32, 1565 (1971).
\bibitem{okada2013observation} Y. Okada \textit{et al.}, Science 341, 1496 (2013).
\bibitem{Corso:1995ky} S. D. Corso, B. Liautard, and J. C. Tedenac, Journal of Phase Equilibria 16, 308 (1995).
\bibitem{STRAUSS:1967dx} A. J. Strauss, Physical Review 157, 608 (1967).
\bibitem{Fu:2011ia} L. Fu, Physical Review Letters 106, 106802 (2011).
\bibitem{Xu:2012bma} S.-Y. Xu, C. Liu, N. Alidoust, and M. Neupane, Nature 3, 1192 (2012).
\bibitem{Dziawa:2012hx} P. Dziawa \textit{et al.}, Nature Materials 11, 1 (2012).
\bibitem{Shu:2015hk} G. J. Shu, S. C. Liou, S. Karna, R. Sankar, M. Hayashi, M.-W. Chu, and F. C. Chou, Applied Physics Letters 106, 122101
(2015).
\bibitem{Xu:2015kx} S.-Y. Xu \textit{et al.}, SCIENCE 347, 294 (2015).
\bibitem{liu2014discovery} Z. Liu \textit{et al.}, Science 343,
864 (2014).
\bibitem{neupane2014observation} M. Neupane \textit{et al.}, Nature communications 5, 3786 (2014).
\bibitem{Wang:cmANUFtU} Z. Wang, Y. Sun, X.-Q. Chen, C. Franchini, G. Xu, H. Weng, X. Dai, and Z. Fang, Physical Review B 85, 195320 (2012).
\bibitem{Steigmann:1968kv} G. A. Steigmann and J. Goodyear, 24, 1062 (1968).
\bibitem{liang2015ultrahigh} T. Liang, Q. Gibson, M. N. Ali, M. Liu, R. Cava, and N. Ong, Nature materials 14, 280 (2015).
\bibitem{Guo:2016di} S.-T. Guo, R. Sankar, Y.-Y. Chien, T.-R. Chang, H.-T. Jeng, G.-Y. Guo, F. C. Chou, and W.-L. Lee, Scienticfic Reports 6, 27487 (2016).
\bibitem{Arushanov:1980u} E. K. Arushanov, Progress in Crystal Growth and Characterization 3, 211 (1980).
\bibitem{Ali:2014kc} M. N. Ali, Q. Gibson, S. Jeon, B. B. Zhou, A. Yazdani, and R. J. Cava, Inorganic Chemistry 53, 4062 (2014).
\bibitem{burkov2011aa} A. Burkov, Phys. Rev. Lett. 107, 127205 (2011).
\bibitem{Hills:vz} R. D. Y. Hills, A. Kusmartseva, and F. V. Kusmartsev, Physical Review B 95, 214103 (2017).
\bibitem{Wan:2011hi} X. Wan, A. M. Turner, A. Vishwanath, and S. Y. Savrasov, Physical Review B 83, 205101 (2011).
\bibitem{Xu:2015bt} S.-Y. Xu \textit{et al.}, Science Advances 1, e1501092 (2015).
\bibitem{Okada:2015jqa} S.-Y. Xu \textit{et al.}, Science 349, 613 (2015).
\bibitem{yang2015weyl} L. Yang \textit{et al.}, Nature physics 11, 728 (2015).
\bibitem{Huang:2015ic} S.-M. Huang \textit{et al.}, Nature Communications 6, 7373 (2015).
\bibitem{Boller:1963jja} H. Boller and E. Parthe, Acta Crystallographica 16, 1095 (1963).
\bibitem{Neupane:2016cza} M. Neupane \textit{et al.}, Physical Review B 93, 201104 (2016).
\bibitem{Bian:2016bn} G. Bian \textit{et al.}, Nature Communications 7, 1 (2016).
\bibitem{Sankar:2017kqa} R. Sankar, G. N. Rao, I. P. Muthuselvam, T.-R. Chang, H. T. Jeng, G. S. Murugan, W.-L. Lee, and F. C. Chou, Journal of
Physics: Condensed Matter 29, 095601 (2017).
\bibitem{KleinHaneveld:1964bv} A. J. Klein Haneveld and F. Jellinek, Recueil des Travaux Chimiques des Pays-Bas 83, 776 (1964).
\bibitem{Pauling:1968te} L. Pauling, Proceedings of the National Academy of Sciences of the United States of America 60, 59 (1968).
\bibitem{Anderson:1987iia} P. W. Anderson, Science 235, 1196 (1987).
\bibitem{Su:2018vz} C. C. Su, C. S. Li, T. C. Wang, S. Y. Guan, R. Sankar, F. C. Chou, C. S. Chang, W. L. Lee, G. Y. Guo, and T. M. Chuang,
New J. Phys., to be published (2018).
\bibitem{Yang:2018im} S.-Y. Yang, H. Yang, E. Derunova, S. S. P. Parkin, B. Yan, and M. N. Ali, Advances in Physics: X 3, 1414631 (2018).

\end{thebibliography}

\end{document}